\documentstyle[rmp,harvard,eqsecnum,aps,epsf]{revtex}
\citationstyle{dcu}
\citationmode{abbr}
\def\gtap{\ \raisebox{-.4ex}{\rlap{$\sim$}} \raisebox{.4ex}{$>$}\ }
\def\eq#1{{eq.~(\ref{#1})}}
\def\eqs#1#2{{eqs.~(\ref{#1})--(\ref{#2})}}

\let\vev\VEV
\def\abs#1{\left| #1\right|}
\def\mod#1{\abs{#1}}
\def\Im{\mathop{\mbox{Im} }}
\def\Re{\mathop{\mbox{Re} }}
\def\Tr{\mathop{\mbox{Tr} }}

\def\qq{$\vev{\bar q q}$ }
\def\GG{$\vev{\alpha_s GG/ \pi}$ }
\def\eps{$\varepsilon$ }
\def\e{$\varepsilon'$ }
\def\ee{$\varepsilon'/\varepsilon$ }
\def\CP{$CP$ }
\newcommand{\bea}{\begin{eqnarray}}
\newcommand{\beq}{\begin{equation}}
\newcommand{\eea}{\end{eqnarray}}
\newcommand{\eeq}{\end{equation}}
\newcommand{\nnu}{\nonumber}

\begin{document}


\title{Estimating $\varepsilon'/\varepsilon$. A Review\\[-2 ex] }

\author{\large Stefano Bertolini and Marco Fabbrichesi}

\address{INFN, Sezione di Trieste and\\ 
Scuola Internazionale Superiore di Studi Avanzati, I-34013 Trieste, Italy\\}

\author{\large Jan O. Eeg}

\address{Fysisk Institutt, Universitetet i Oslo, N-0316 Oslo, Norway.\\}

\maketitle


\begin{abstract}
The real part of \ee measures 
direct \CP violation in the decays of the neutral kaons in two 
pions. It is a fundamental quantity which has justly attracted a great
deal of theoretical as well as experimental work. Its
determination may answer the question of whether
\CP violation is present only 
in the mass matrix of neutral kaons (the superweak scenario) or  
 also at work directly in the decays. 
After a brief
historical summary, we discuss the present and 
expected experimental sensitivities.
In the light of these, we come to the problem of estimating \ee in the
standard model.  
We review the present (circa 1998) status of the theoretical
predictions of $\varepsilon'/\varepsilon$.
The short-distance part of the computation is now known to the
next-to-leading order in QCD and QED and therefore
well under control. On the other
hand, the evaluation of the hadronic matrix elements of the relevant operators
is where most of the theoretical uncertainty still resides. 
We analyze the results
of the currently most developed calculations.
The values of the
$B_i$ parameters in the various approaches are discussed,
together with the allowed range of the relevant combination of the
Cabibbo-Kobayashi-Maskawa entries 
Im $V_{td}V^*_{ts}$.
We conclude by summarizing and comparing all
up-to-date predictions of $\varepsilon'/\varepsilon$. 
Because of the intrinsic uncertainties of 
the long-distance computations, values ranging from
$10^{-4}$ to a few times $10^{-3}$ can be accounted for in the standard model.
Since this range covers most of the present experimental
uncertainty, it is unlikely that new physics effects
can be disentangled from the standard model prediction.
For updates on the review and additional material see 
{\tt http://www.he.sissa.it/review/}.
\end{abstract}



\tableofcontents


\section{What \ee Is and Why It Is Important to Know Its Value}

A \CP transformation consists in
a parity ($P$) flip followed by charge conjugation ($C$). 
It was promoted~\cite{Landau:1957tp} to
a symmetry of nature after parity 
was shown to be maximally violated
in weak interactions~\cite{Wu:1957my}.

Until 1963 the \CP symmetry was thought to be exactly conserved in all
physical processes. That year, J.M. Christenson, J.W. Cronin, V.L. Fitch and 
R. Turlay~\citeyear{Christenson:1964fg} announced the surprising result that
the \CP symmetry was indeed violated in hadronic decays of the neutral kaons.

In order to interpret the experimental evidence we must consider
the strong Hamiltonian eigenstates $K^0$ and its \CP conjugate $\bar K^0$ 
as an admixture of the physical short-lived $K_S$ component---which decays 
predominantly into two 
pions---and the physical long-lived $K_L$ component---which decays 
predominantly semileptonically and into three pions.

The two and three pion final states are, respectively,
even and odd under a \CP transformation.
Therefore, in the absence of \CP
violating interactions, we would expect the $K_{S,L}$ mass eigenstates
to coincide with the states
\bea
K_1 &=& (K^0 +\bar K^0)/\sqrt{2}
\nnu\\
K_2 &=& (K^0 -\bar K^0)/\sqrt{2}\ ,
\eea
which exhibit a definite \CP parity, even and odd respectively
(we choose $CP\; |K^0 \rangle = |\bar K^0 \rangle$). 

What was observed in 1963 was that also
$K_L$ decays a few times in a thousand 
into a two-pion final state, and accordingly that the
\CP symmetry is not exact.

The violation of \CP in $K_{S,L}$ decays can proceed indirectly,
via a mismatch between the \CP eigenstates $K^0_{1,2}$
and the weak mass eigenstates $K_{S,L}$ introduced by a $CP$-violating
impurity in the $\bar K^0$-$K^0$ mixing, 
and/or directly in the decays of the \CP eigenstates. 
Both effects are 
usually parameterized in terms of the ratios (for a recent theoretical
review see, for instance, \cite{deRafael:1994xx})
\beq
\eta_{00} \equiv 
\frac{\langle \pi^0 \pi^0 | {\cal L}_W | K_L \rangle}
{\langle \pi^0 \pi^0 | {\cal L}_W | K_S \rangle}
\label{eta00}
\eeq
and
\beq
\eta_{+-} \equiv \frac{\langle \pi^+ \pi^- | {\cal L}_W | K_L \rangle}
{\langle \pi^+ \pi^- | {\cal L}_W | K_S \rangle} \ , 
\label{eta+-}
\eeq
where ${\cal L}_W$ represents the $\Delta S = 1$ weak lagrangian.
Eqs.~(\ref{eta00}) and (\ref{eta+-})
can be written as
\bea
\eta_{00}& =& \varepsilon - \frac{2 \varepsilon'}{1 - \omega \sqrt{2}}
 \simeq \varepsilon - 2 \varepsilon' \; ,
\nnu \\
\eta_{+-} & =&  \varepsilon + \frac{\varepsilon'}{1 + \omega/ \sqrt{2}}
\simeq  \varepsilon + \varepsilon' \; ,
\label{eta00eta+-}
\eea
where the complex parameters $\varepsilon$ and $\varepsilon'$
parameterize indirect (via $K_1$-$K_2$ mixing) and direct (in the $K_1$ 
and $K_2$ decays) \CP violation respectively.
The $K_{S,L}$ eigenstates are given by
\bea
K_S &=& \frac{K_1 + \bar\varepsilon\ K_2}{\sqrt{1+|\bar\varepsilon|^2}}
\nnu\\
K_L &=& \frac{K_2 + \bar\varepsilon\ K_1}{\sqrt{1+|\bar\varepsilon|^2}}\; ,
\label{KL-KS}
\eea
where $\bar\varepsilon$ is a (complex) parameter 
of order $10^{-3}$ which depends on the chosen
\CP phase convention. The $K_1 - K_2$ mixing parameter $\bar\varepsilon$
is simply related to the observable parameter $\varepsilon$ in 
\eq{eta00eta+-} (see \eq{epsbareps} below).
The parameter $\omega$ measures the ratio:
\beq
|\omega| \equiv 
\left|\frac{\langle ( \pi \pi )_{(I=2)} | {\cal L}_W | K_S \rangle}
{\langle ( \pi \pi )_{(I=0)} | {\cal L}_W | K_S \rangle}\right|  
\simeq 1/22.2 \, ,
\label{omega}
\eeq
where $I=1$ and 2 stand for the isospin states of the final pions. 
For notational convenience, 
we identify in the following $\omega$ with its absolute value. 
The smallness of the experimental value
of $\omega$ given by~(\ref{omega}) is known as
the $\Delta I = 1/2$ selection rule of $K\to \pi\pi$ 
decays~\cite{Gell-Mann:1954aa}.

In terms of the $K_{S,L}$ decay amplitudes,
the \CP violating parameters \eps and \e are given by
\beq
\varepsilon =
\frac{\langle ( \pi \pi )_{(I=0)} | {\cal L}_W | K_L \rangle}
{\langle ( \pi \pi )_{(I=0)} | {\cal L}_W | K_S \rangle} \, , 
\label{eps}
\eeq
 and
\beq
\varepsilon' = 
\frac{\varepsilon}{\sqrt{2}} \left\{
\frac{\langle ( \pi \pi )_{I=2} | {\cal L}_W | K_L \rangle}
{\langle ( \pi \pi )_{I=0} | {\cal L}_W | K_L \rangle}
 - \frac{\langle ( \pi \pi )_{I=2} | {\cal L}_W | K_S \rangle}
{\langle ( \pi \pi )_{I=0} | {\cal L}_W | K_S \rangle} \right\} 
\label{eps'} \ .
\eeq
 From \eqs{KL-KS}{eps'} one sees that direct \CP violation arises due 
to the relative misalignment of the $K_S$ and $K_L$ $I=0,2$ amplitudes
and it is suppressed by the $\Delta I = 1/2$ selection rule.

According to the Watson theorem \cite{Watson:1952ji}, 
we can write the generic amplitudes 
for $K^0$ and $\bar K^0$ to decay into two pions as
\bea
\langle ( \pi \pi )_{(I)} | {\cal L}_W | K^0 \rangle & =& - i A_I 
\, \exp \: (i \, \delta_I)  \nnu \\
\langle ( \pi \pi )_{(I)} | {\cal L}_W | \bar K^0 \rangle & =& - i  A_I^* 
\, \exp \: (i \, \delta_I) \; ,
\label{def2}
\eea
where the phases $\delta_I$ arise from the pion final-state 
interactions (FSI).  Using \eq{def2} and the approximations
\beq
|\bar\varepsilon|\Im A_0 \ll \Re A_0\ , \quad\quad |\bar\varepsilon|^2 \ll 1
\label{approx1}
\eeq
the $\varepsilon'$ parameter in \eq{eps'}
can be written as
\beq
\varepsilon' =
e^{i (\pi/2 + \delta_2 - \delta_0)} \: \frac{\omega}{\sqrt{2}}
\left( \frac{{\rm Im} A_2}{{\rm Re} A_2} - 
\frac{{\rm Im} A_0}{{\rm Re} A_0} \right) \, ,
\label{defeps'}
\eeq
where the parameter $\omega$ can be written as
\beq
\omega = \frac{\Re A_2}{\Re A_0}\ .
\label{defomega}
\eeq

By decomposing the $\Delta S=2$ weak lagrangian for the $\bar K^0$-$K^0$ 
system in a dispersive and an absorptive components as 
$M - i\ \Gamma/2$, where $M$ and $\Gamma$
are $2\times 2$ hermitian matrices ($CPT$ symmetry is assumed), 
one obtains for $\varepsilon$ the expression
\beq
\varepsilon =
\sin\theta_\epsilon \: e^{i \theta_\epsilon}
\left(\frac{{\rm Im} M_{12}}{\Delta M_{LS}}
+ \frac{{\rm Im} A_0}{{\rm Re} A_0} \right) \, , 
\label{defeps}
\eeq
where $\Delta M_{LS}$ is the mass difference of the $K_L$-$K_S$ mass
 eigenstates, $M_{12}$ is the $K_1-K_2$ entry in the mass matrix, and
\beq
\theta_\epsilon = \tan^{-1}\left(2 \Delta M_{LS}/ \Delta\Gamma_{SL}\right)
\simeq \pi/4  \; .
\label{thetaeps}
\eeq
In obtaining \eq{defeps}, in addition to the approximations of \eq{approx1},
the experimental observations that $\Delta M_{LS}\simeq \Gamma_S/2$ and 
$\Gamma_L \ll \Gamma_S$ have been used. With the above approximations
one also obtains a simple relation between the observable parameter
$\varepsilon$ and the phase-convention dependent parameter $\bar\varepsilon$,
\beq
\varepsilon = \bar\varepsilon + i\ \frac{\Im A_0}{\Re A_0}\ .
\label{epsbareps}
\eeq
For detailed discussions on the role and implications of the phase 
conventions we refer the reader to the reviews of 
\cite{Chau:1983da} and \cite{Nir:1992uv}.

It is useful to bear
in mind that the real and imaginary parts of $A_{0,2}$ are always
taken with respect to the $CP$-violating phase and not 
the final-state strong interaction phases
that have  already been extracted in \eq{def2}.
A simpler form of \eq{defeps'}, in which ${\rm Im} A_0 =0$, is found in
those papers that follow the Wu-Yang phase convention. In this case
$\varepsilon = \bar\varepsilon$.

In the standard model, \e can be in principle different from zero
because the $3 \times 3$ 
Cabibbo-Kobayashi-Maskawa (CKM) 
matrix $V_{ij}$, which appear in the weak charged currents
of the quark mass eigenstate,
can be in general complex~\cite{Kobayashi:1973pm}:
\beq
\left( \begin{array}{c c c} 
V_{ud} & V_{us}  & V_{ub}  \nnu \\
V_{cd} & V_{cs}  & V_{cb}  \nnu \\
V_{td} & V_{ts}  & V_{tb} 
\end{array} \right)  \approx
\left( \begin{array}{c c c} 
1 - \lambda^2/2 & \lambda & A \lambda^3 ( \rho - i \eta) \nnu \\
-\lambda - i A^2 \lambda^5 \eta & 1 - \lambda^2/2 & A \lambda^2 \nnu \\
A \lambda^3 (1 - \rho - i \eta) & - A \lambda^2 (1 + i \lambda^2 \eta) & 1
\end{array} \right) \, . \label{KM}
\eeq
In \eq{KM} we have used the Wolfenstein parameterization in terms of four
parameters: $\lambda$, $A$, $\eta$ and $\rho$ and retained all imaginary
terms for which unitarity is achieved up to $O(\lambda^5)$, with $\lambda =
|V_{us}| = 0.22$.
On the other
hand, in other models like the superweak theory~\cite{Wolfenstein:1964ks},
the only source of \CP violation resides in
the  $K^0$-$\bar K^0$ oscillation, and \e vanishes. It is therefore of great
importance to establish the experimental value of \e and discuss
its theoretical predictions within the standard model and beyond.

\subsubsection{A Brief History}

The presence in nature of indirect \CP violation
is an experimentally well established result~\cite{Barnett:1996hr}
\beq
|\varepsilon| = (2.266 \pm 0.017) \times 10^{-3} \label{eps_K}\ , 
\eeq
which can be understood both qualitatively and quantitatively 
in the framework of the standard model of electroweak interactions
with three generations of quarks. 
On the other hand, after 34 years from the discovery of Christenson et al. 
there is still no conclusive
experimental evidence for a non-vanishing $\varepsilon'$.

The ratio \ee is measured from 
\beq
\left|\frac{\eta_{+-}}{\eta_{00}}\right|^2 \simeq 
1 + 6 \; \Re\: \frac{\varepsilon '}{\varepsilon} \; .
\eeq
As discussed above, a non-vanishing $\varepsilon'/\varepsilon$
gives the experimental evidence for direct \CP violation. Due to the
accuracy in the counting of $K_{L,S}$ decays required by the 
expected smallness of $|\varepsilon'/\varepsilon|$ in the standard model
its detection represents a hard experimental challenge.
\begin{figure}
\epsfxsize=9cm
\centerline{\epsfbox{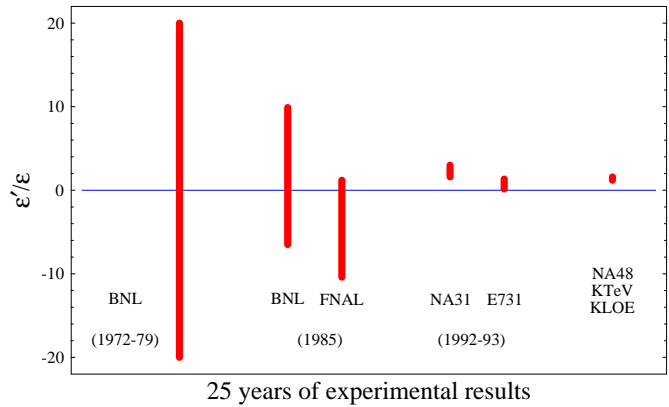}}
\caption{25 years of experiments on \ee (in units of $10^{-3}$).
 The last mark on the right, 
at the average central value of the 1992-93
experiments, shows the experimental precision expected in the forthcoming
years.}
\label{storia}
\end{figure}

As it shown in Fig. \ref{storia}, the experimental
error in the determination 
of this quantity has been dramatically reduced over the
years from
$10^{-2}$ in the 
70's~\cite{Holder:1972aa,Banner:1972aa,Christenson:1979tu,Christenson:1979tt}
to $3.5 \times 10^{-3}$ in 1985~\cite{Black:1985vj,Bernstein:1985vk} and to
roughly 
$7 \times 10^{-4}$ in the last run of experiments in 1992 at CERN and
FNAL 
that obtained respectively~\cite{Barr:1993rx,Gibbons:1997fw}
\bea
\Re \: \varepsilon '/\varepsilon & = &
(23 \pm 3.6 \pm 5.4)  \times   10^{-4} \; \mbox{(NA31)} \; , 
\label{NA31}\\
\Re \: \varepsilon '/\varepsilon & = &
(7.4 \pm 5.2 \pm 2.9)  \times   10^{-4} \; \mbox{(E731)} \; ,
\label{E731}
\eea
where the first error is statistical and the second one systematic.
As the
reader can see, the agreement between the two experiments is
within two standard deviations.
Moreover, only the CERN result is definitely different from zero.

Before the end of 1999 the new FNAL (E832-KTeV)~\cite{Odell:1997ktev} and 
CERN (NA48)~\cite{Holder:1997na48} experiments
should provide data with a precision of $(1 \div 2) \times 10^{-4}$ and 
hopefully settle the issue of whether \ee is or is not zero. Results 
of the same precision should also be achieved at 
DA$\Phi$NE (KLOE)~\cite{Patera:1997dafne}, the Frascati $\Phi$-factory.
For a detailed account of the experimental setups and a critical discussion
of the issues involved see the review article by~\cite{Winstein:1993sx}.

From the theoretical point of view, the prediction of the
value of \ee has gone
through almost twenty years of increasingly more accurate analyses.
By the end of the 70's, it had been recognized
that within the standard
model with three generations of quarks, direct \CP violation is natural
and therefore the model itself is distinguishable from the superweak model.
This understanding was the result of an intensive work leading to
the identification of the dominant
operators responsible for the transition, the so-called penguin operators,
and the role played by QCD in their 
generation~\cite{Vainshtein:1975sv,Vainshtein:1977sc}.
Typical estimates during this period gave \ee  
$\sim 10^{-3}-10^{-2}$~\cite{Ellis:1976fn,Ellis:1977uk,Gilman:1979bc,Gilman:1979wm}.

The next step came in the  80's as the gluon penguin
operators above were
joined by the electromagnetic operators together with other isospin breaking
corrections~\cite
{Bijnens:1984ye,Donoghue:1986nm,Buras:1987wc,Lusignoli:1989fz}. 
It was then recognized that these contributions tend to 
make \e smaller because they have the opposite sign compared to the gluonic
penguin contributions.
This part of the computation became particularly
critical when by the end of the decade
 it was realized that the increasingly large mass of the $t$
quark would lead to an increasingly large contribution of the electroweak 
penguins~\cite{Flynn:1989iu,Buchalla:1990we,Paschos:1991as,Lusignoli:1992bm}. 
This meant a potentially vanishing value for \ee because of the destructive
interference between the two contributions.

By the 90's the entire subject was mature for a systematic exploration as
the short-distance part was brought under control by the 
next-to-leading order (NLO) determination
of the Wilson coefficients of all relevant 
operators~\cite{Buras:1992jm,Buras:1993zv,Buras:1993tc,Buras:1993dy,Ciuchini:1993tj,Ciuchini:1994vr}. This theoretical achievement together with
the discovery of the $t$ quark
(and the determination of its mass~\cite{Barnett:1996hr}) 
removed two of the 
largest sources of uncertainty in the prediction. At the same
time, independent efforts
were brought to bear on the matrix elements 
estimate. All combined improvements made possible the
current predictions of the value of
 \ee within the standard 
model~\cite{Heinrich:1992en,Paschos:1996,Buras:1993dy,Buras:1996dq,Ciuchini:1993tj,Ciuchini:1995cd,Ciuchini:1997kd,Bertolini:1996tp,Bertolini:1997nf} that
we are to going to review.

\subsubsection{Outline}

The analysis of \ee can be divided into the short-distance (perturbative)
part
and the long-distance (mainly non-perturbative) part. 
As already mentioned, the  short-distance part is
by now known at the NLO level and is therefore  under control. 
This part of the computation is briefly reviewed 
in the next section. The  long-distance component has been studied by
a variety of approaches---lattice QCD, 
phenomenological estimates and QCD-like models---all of which
are  eventually combined with
 chiral perturbation theory. As the  long-distance
 part is the most uncertain, we will 
spend most of the review on that issue. Section II and III set the common
ground on which all approaches are based. Section IV reviews the
various determinations of the hadronic matrix elements. After a brief 
detour, in section V, to determine the relevant CKM matrix 
elements, in sections VI and VII 
we bring all elements together to discuss some simple models.
We then
 summarize the current theoretical predictions in
the standard model and comment on the issue of new physics. 

For a broader view on \CP violation which complements the present review,
especially in the attention to the experimental issues,
the reader is encouraged to consult the article
previously published in this journal~\cite{Winstein:1993sx}.

\section{The Quark Effective Lagrangian and the NLO Wilson Coefficients}

The study of kaon decays within the standard model is made complicated by the
huge scale differences involved. Energies as far apart as the mass of
the $t$ quark and the mass of the pion must be included. The most satisfactory
framework for dealing with physical systems defined across
 different energy scales is
that of effective theories~\cite{Weinberg:1980wa,Georgi:1984zz}.  
The transition amplitudes of an effective theory are
assumed to be factorizable in high- and low-energy parts.
The degrees of freedom
at the higher scales are step-by-step integrated out, retaining
only the effective operators made of the lighter degrees of freedom. 
The short-distance physics, obtained from  integrating out 
the heavy scales, is encoded in the
Wilson coefficients that multiply the effective operators. Their evolution with
the energy scale is described by the 
renormalization group equations~\cite{Wilson:1971ag}.

\begin{figure}
\epsfxsize=8cm
\centerline{\epsfbox{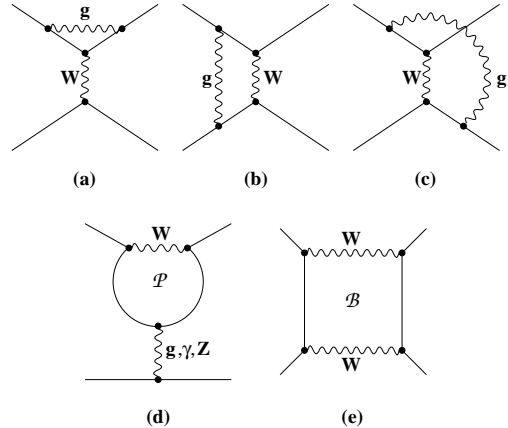}}
\caption{Standard model contributions to the matching of the
quark operators in the effective flavor-changing
Lagrangian.}
\label{fulldiag}
\end{figure}

Figure \ref{fulldiag} shows the typical diagrams that  in the standard
model generate the operators of the effective $\Delta S=1$ Lagrangian.

The $\Delta S=1$
quark effective lagrangian at a scale $\mu < m_c$ can be 
written~\cite{Shifman:1977tn,Gilman:1979bc,Bijnens:1984ye,Lusignoli:1989fz}
as
 \beq
 {\cal L}_{W}= - \sum_i  C_i(\mu) \; Q_i (\mu) \; ,
 \label{Lquark}
\eeq
where 
\bea
C_i(\mu) =  \frac{G_F}{\sqrt{2}} V_{ud}\,V^*_{us} 
 \Bigl[z_i(\mu) + \tau\ y_i(\mu) \Bigr] \; .
\label{Lqcoef}
\eea
Here $G_F$ if the Fermi coupling, the functions $z_i(\mu)$ and
 $y_i(\mu)$ are the
 Wilson coefficients and $V_{ij}$ the
CKM matrix elements; $\tau = - V_{td}
V_{ts}^{*}/V_{ud} 
V_{us}^{*}$. 
According to the standard parameterization of the CKM matrix, in order to
determine $\varepsilon'/\varepsilon$, we only
need to consider the $y_i(\mu)$ components, 
which control the $CP$-violating part of the Lagrangian.
The coefficients   $y_i(\mu)$, and $z_i(\mu)$ contains all the dependence
of short-distance
 physics, and depend on the $t,W,b,c$ masses, the intrinsic QCD scale
$\Lambda_{\rm QCD}$ and the renormalization scale $\mu$.

The $Q_i$ are the effective four-quark operators obtained 
in the standard model by integrating out 
the vector bosons and the heavy quarks $t,\,b$ and $c$. A convenient
and by now standard
basis includes the following ten  operators:
 \beq
\begin{array}{rcl}
Q_{1} & = & \left( \overline{s}_{\alpha} u_{\beta}  \right)_{\rm V-A}
            \left( \overline{u}_{\beta}  d_{\alpha} \right)_{\rm V-A}
\, , \\[1ex]
Q_{2} & = & \left( \overline{s} u \right)_{\rm V-A}
            \left( \overline{u} d \right)_{\rm V-A}
\, , \\[1ex]
Q_{3,5} & = & \left( \overline{s} d \right)_{\rm V-A}
   \sum_{q} \left( \overline{q} q \right)_{\rm V\mp A}
\, , \\[1ex]
Q_{4,6} & = & \left( \overline{s}_{\alpha} d_{\beta}  \right)_{\rm V-A}
   \sum_{q} ( \overline{q}_{\beta}  q_{\alpha} )_{\rm V\mp A}
\, , \\[1ex]
Q_{7,9} & = & \frac{3}{2} \left( \overline{s} d \right)_{\rm V-A}
         \sum_{q} \hat{e}_q \left( \overline{q} q \right)_{\rm V\pm A}
\, , \\[1ex]
Q_{8,10} & = & \frac{3}{2} \left( \overline{s}_{\alpha} 
                                                 d_{\beta} \right)_{\rm V-A}
     \sum_{q} \hat{e}_q ( \overline{q}_{\beta}  q_{\alpha})_{\rm V\pm A}
\, , 
\end{array}  
\label{Q1-10} 
\eeq
where $\alpha$, $\beta$ denote color indices ($\alpha,\beta
=1,\ldots,N_c$) and $\hat{e}_q$  are the quark charges 
($\hat{e}_u = 2/3$,  $\hat{e}_d=\hat{e}_s =-1/3$). Color
indices for the color singlet operators are omitted. 
The labels \mbox{$(V\pm A)$} refer to the Dirac structure
\mbox{$\gamma_{\mu} (1 \pm \gamma_5)$}.

The various operators originate from different
diagrams of the fundamental theory. 
First, at the tree level, we only have the current-current operator
$Q_2$ induced by $W$-exchange. Switching on QCD, a one-loop correction
to $W$-exchange (like in Fig.~\ref{fulldiag}b,c) will induce $Q_1$.
 Furthermore, QCD
through the penguin loop (Fig.~\ref{fulldiag}d) induces the gluon
penguin operators
 $Q_{3-6}$.
The gluon penguin contribution is split in four
components because of the splitting of the gluonic
coupling into a right- and a left-handed part 
and the use of the $SU(N_c)$ relation
\beq
2\ T^a_{\alpha\delta} T^a_{\gamma\beta} =
\delta_{\alpha\beta}\delta_{\gamma\delta} -
\frac{1}{N_c}\delta_{\alpha\delta}\delta_{\gamma\beta} \ ,
\label{colorfierz}
\eeq 
where $N_c$ is the number of colors,
$a= 1,\ ...,\ N_c^2 - 1$ and $T^a$ are the properly normalized
$SU(N_c)$ generators, $\Tr T^a T^b = 1/2\ \delta^{ab}$, 
in the fundamental representation.  
Electroweak loop diagrams---where
the penguin gluon is replaced by a photon or a $Z$-boson and also box-like
diagrams---induce $Q_{7,9}$ and also a part of $Q_3$. The operators $Q_{8,10}$
are induced by the QCD renormalization of the electroweak loop
operators $Q_{7,9}$.

Even though  the operators in \eq{Q1-10} are not all independent, this basis
is of particular interest for any 
numerical analysis because it is that employed
for the calculation of the Wilson coefficients 
to the NLO order in $\alpha_s$ and 
$\alpha_e$~\cite{Buras:1992jm,Buras:1993zv,Buras:1993tc,Buras:1993dy,Ciuchini:1993tj,Ciuchini:1994vr} and we will use it throughout.

\begin{figure}
\epsfxsize=8cm
\centerline{\epsfbox{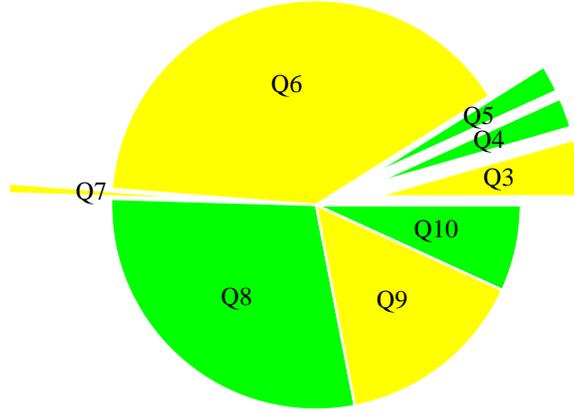}}
\caption{Relative contributions to \ee of the operators in \eq{Q1-10}. 
Operators giving a (negative)
positive contribution are depicted in (dark) light gray. All matrix elements
are taken in the vacuum saturation approximation. The operators $Q_8$
and $Q_{10}$ are generated by QCD and do not receive contributions
from the one-loop matching. }
\label{pie}
\end{figure}

Anticipating our discussion,
the pie chart in Fig.~\ref{pie} 
 shows pictorially  the relative importance of the operators
in \eq{Q1-10} in the final determination of the value of \ee, as obtained in
the vacuum saturation approximation to the hadronic matrix elements. 
In particular, Fig. \ref{pie} shows the crucial competition between gluonic
and electroweak penguins in the determination of the value of
\ee. Such a destructive interference might accidentally lead to a vanishing
\ee even in the presence of a source of direct $CP$ violation.
This feature adds to the theoretical challenge of predicting \ee
with the required accuracy.

While there exist other possible operators in addition to those listed
in eq.~(\ref{Q1-10}), they
are numerically irrelevant within the
standard model. For instance, the two
 operators
\beq 
Q_{11}  =  
\frac{g_s}{8 \pi^2} \: \bar s \: \left[ 
m_d R + m_s L \right] 
\: \sigma \cdot G \: d 
\;  \quad \quad \mbox{and} \quad \quad
Q_{12} =  
\frac{e}{8 \pi^2} \: \bar s \: \left[ 
m_d R + m_s L \right] 
\: \sigma \cdot F \: d \; , 
\eeq
where $R =  \left(1 + \gamma_5 \right)/2$ and 
$L = \left(1 - \gamma_5 \right)/2$,
are present. These operators are induced by gluon and photon penguins 
with a free gluon (photon) leg. 
The matrix elements of these operators give a vanishingly small 
contribution to $K\to \pi\pi$ 
decays~\cite{Bertolini:1995qk,Bertolini:1997ir}.

In table \ref{Wcoefs} we summarize in a synthetic way the diagrammatic origin
of the contributions
to the various Wilson coefficients when considering the 
one-loop matching of the quark effective
lagrangian with the full electroweak theory.

\vbox{
\begin{table}[thb]
\caption[]{Contributions to the one-loop matching of the
$\Delta S=1$ Wilson coefficients at $\mu=m_W$. 
The notation refers to that of \eq{Lquark} and Fig. \ref{fulldiag}.
Non-vanishing contributions to $C_8$ and $C_{10}$ arise via the
QCD renormalization of the operators $Q_7$ and $Q_9$, respectively.
}
\begin{center}
\begin{tabular}{c c c c c c c c c c c}
$\mu=m_W$   &$C_1$&$C_2$&$C_3$&$C_4$&$C_5$&$C_6$&$C_7$&$C_8$&$C_9$&$C_{10}$\\
\hline
Tree             & &$\surd$& & & & & & & & \\ 
Tree + $g$       &$\surd$&$\surd$& & & & & & & & \\ 
Tree + $\gamma$  & &$\surd$& & & & & & & & \\ 
${\cal P}_g$     & & &$\surd$&$\surd$&$\surd$&$\surd$ & & & & \\ 
${\cal P}_\gamma$& & & & & & &$\surd$& &$\surd$& \\ 
${\cal P}_Z$     & & &$\surd$& & & &$\surd$& &$\surd$ & \\
${\cal B}$       & & &$\surd$& & & & & &$\surd$ & \\
\end{tabular}
\label{Wcoefs}
\end{center}
\end{table}}

Having established the operator basis, a full two-loop calculation
(up to $\alpha_s^2$ and $\alpha_s \alpha_{em}$) of the quark operator
anomalous dimensions is performed. This
calculation allows us via renormalization group methods
to evaluate the Wilson coefficients at the typical
scale of the process, thus resumming (perturbatively)
potentially large logarithmic effects to a few 10\% uncertainty.
As already mentioned, the size of the Wilson coefficients
 at the hadronic scale (of the order of 1 GeV) 
depends on $\alpha_s$ and the threshold masses $m_t$, $m_W$, $m_b$ and $m_c$. 
The top quark mass dependence enters in the penguin coefficients 
$y_i(\mu)$ via the initial matching conditions for the renormalization group
equations.

Small differences in the short-distance  input parameters are
present in the various treatment in the literature. 
In order to give the reader an idea of the
ranges used, we list below some of the values.

The most  
recent determination of the running strong coupling
in the $\overline{MS}$ scheme is~\cite{Barnett:1996hr}
\beq
\alpha_s (m_Z) = 0.119 \pm 0.002 \, ,
\eeq
which at the NLO corresponds to
\beq
\Lambda^{(4)}_{\rm QCD} = 340 \pm 40 \: \mbox{MeV} \, .
\label{lambdone}
\eeq

For $m_t$ we take the value~\cite{Tipton:1996aa}
\beq
m_t^{\rm pole} = 175 \pm 6 \:\:\mbox{GeV} \, . 
\label{mtpole}
\eeq
The knowledge of the top quark mass is one important ingredient in the
reduced uncertainty of the recent estimates of \ee.

The relation between the pole mass $M$ and the $\overline{\rm MS}$ 
running mass $m(\mu)$ 
is given at one loop in QCD by 
\beq
m(M)\ =\ 
   {M(q^2=M^2)}\ {\left[ 1-\frac{4}{3}\frac{\alpha_s(M)}{\pi} \right] }
\label{polemass} \ \ ,
\eeq
For the running top quark mass, in the range of $\alpha_s$
considered, we then obtain
\beq
 m_t(m_t)  \simeq 167 \pm 6 \:\:\mbox{GeV} \; ,
\label{mt-mt}
\eeq
which, using the one-loop running, corresponds to 
\beq
 m_t(m_W)  \simeq 177 \pm 7 \:\:\mbox{GeV} \; ,
\label{mt-mw}
\eeq
which is the value to be used as input at the $m_W$ scale for the 
NLO evolution of the Wilson coefficients. In \eq{mt-mw} we have averaged over
the range of $\Lambda^{(4)}_{\rm QCD}$ given in \eq{lambdone}. 

The use of the running top mass in the initial matching of the
Wilson coefficients softens the matching scale dependence present in the
LO analysis.
By taking $\mu=m_t^{\rm pole}$ as the
starting matching scale in place of $m_W$,
and using correspondingly $m_t(m_t)$,  the  
NLO Wilson coefficients of the electroweak and gluon penguins
at $\mu \simeq  1$ GeV, remain stable up to the 10\% percent level. 

For $m_b$  we have the mass range~\cite{Barnett:1996hr}
\beq
m_b^{\rm pole} = 4.5 \div 4.9 \:\:\mbox{GeV} \; ,
\label{mbpole}
\eeq
which corresponds to
\beq
m_b(m_b) = 4.1 \div 4.5 \:\:\mbox{GeV} \; .
\label{mbrun}
\eeq
Analogously, for $m_c$ one has
\beq
m_c^{\rm pole} = 1.2 \div 1.9 \:\:\mbox{GeV} \; ,
\label{mcpole}
\eeq
which corresponds to
\beq
m_c(m_c) = 1.0 \div 1.6 \:\:\mbox{GeV} \; ,
\label{mcrun}
\eeq
Values within the $\overline{MS}$ ranges have to be used
as the quark thresholds in evolving
the Wilson coefficients down to the low-energy scale where the
matching with the hadronic matrix elements is to be performed.

In choosing the quark mass thresholds
one should bear in mind that varying $m_b^{\rm pole}$ 
within the given range affects the final values
of the Wilson coefficients only at the percent level, while varying the
charm pole mass in the whole range given may affects the real part of the
gluon penguin coefficients up to the 20\% level.
We will take 
$m_b(m_b) = 4.4 \:\:\mbox{GeV}$ and
$m_c(m_c) = 1.4 \:\:\mbox{GeV}$.

\vbox{
\begin{table}
\caption{The $\Delta S = 1$ NLO Wilson coefficients relevant for 
CP violation are given at $\mu=1.0$ GeV for 
$m_t(m_W)=177$~GeV, which corresponds to $m_t^{pole}=175$~GeV,
($\alpha=1/128$). In addition one has $y_{1,2}(\mu) = 0$.}
\begin{center}
\begin{tabular}{c r r r r r r}
$\Lambda_{QCD}^{(4)}$ & \multicolumn{2}{c}{ 300 MeV }
                      & \multicolumn{2}{c}{ 340 MeV } 
                      & \multicolumn{2}{c}{ 380 MeV } \\
\hline 
& HV & NDR & HV & NDR & HV & NDR \\
\hline 
$y_3$&$0.0341 $&$0.0298 $&$0.0378 $&$0.0326 $&$0.0420 $&$0.0356 $ \\
$y_4$&$-0.0558 $&$-0.0530 $&$-0.0597 $&$-0.0564 $&$-0.0639 $&$-0.0597 $ \\
$y_5$&$0.0149 $&$0.000687 $&$0.0160 $&$-0.00204 $&$0.0173 $&$-0.00581 $ \\
$y_6$&$-0.0883 $&$-0.100 $&$-0.0994 $&$-0.115 $&$ -0.113 $&$-0.133 $ \\
$y_7/\alpha$&$-0.0202 $&$-0.0210 $&$-0.0195 $&$-0.0209 $&$-0.0188 $&$-0.0209 $ \\
$y_8/\alpha$&$0.184 $&$0.169 $&$0.209 $&$0.192 $&$0.240 $&$0.220 $ \\
$y_9/\alpha$&$-1.70 $&$-1.70 $&$-1.75 $&$-1.74 $&$-1.80 $&$-1.80 $ \\
$y_{10}/\alpha$&$0.735 $&$0.722 $&$0.806 $&$0.790 $&$0.885 $&$0.867 $ \\
\end{tabular}
\label{numWcoefs}
\end{center}
\end{table}}

In table \ref{numWcoefs} we report the numerical values of the NLO
Wilson coefficients relevant for \CP violation in $\Delta S = 1$
processes. The coefficients $y_i(\mu)$ are given at the scale
$\mu = 1$ GeV and are dependent on the choice of the $\gamma_5$
scheme in dimensional regularization. The values in the table refer 
to two commonly used schemes, namely the naive dimensional regularization
(NDR), in which  $\gamma_5$ anticommutes with the Dirac matrices in
$d$ dimensions, and the t' Hooft-Veltman scheme (HV) \cite{'tHooft:1972fi},
in which they anticommute only in four dimensions.
The latter prescription has been shown to be a consistent formulation
of dimensional regularization in the presence of chiral couplings
\cite{Breitenlohner:1977te}.
A consistent calculation of the hadronic matrix elements should match
the unphysical scale and scheme dependence of the Wilson coefficients 
so as to produce a stable amplitude 
at the given order in perturbation theory. We will return on this issue
in Sect IV when discussing the various approaches to the long distance
part of the calculation.

The case of the $\Delta S=2$ theory is treated along similar lines.
The  effective $\Delta S=2$ quark 
lagrangian at scales $\mu<m_c$ is given by
\beq
{\cal L}_{\Delta S = 2}  = -  C_{2S}(\mu) \;  Q_{S2} (\mu) \; ,
\label{lags2}
\eeq
where
\beq
C_{2S}(\mu)   =  
\frac{G_F^2 m_{W}^2 }{4 \pi^2} \left[ \lambda_c^2 \eta_{1} S(x_c) 
 + \lambda_t^2 \eta_{2} S(x_t)
 + 2 \lambda_c \lambda_t \eta_3 S(x_c , x_t)\right] b(\mu) 
\label{lags2C}
\eeq
where $\lambda_j = V_{jd} V_{js}^{*}$,
$x_i = m_i^2 / m_W^2$.
We denote by
$Q_{S2}$ the $\Delta S=2$ local four quark operator
\beq
Q_{S2} =(\bar{s}_L \gamma^{\mu} d_L) (\bar{s}_L \gamma_{\mu} d_L)
\, , \label{QS2}
\eeq
which is the only local operator of dimension six in the standard model.

The integration of the electroweak loops leads to the
 Inami-Lim functions~\cite{Inami:1981fz} $S(x)$ and $S(x_c, x_t)$, the exact
expressions of which can be found in the reference quoted, 
depend on the masses of the charm and top quarks and describe the 
$\Delta S = 2$ transition amplitude in the absence of strong interactions.

The short-distance QCD corrections are encoded in the coefficients $\eta_1$,
$\eta_2$ and $\eta_3$ with a common scale-dependent
factor $b(\mu)$ factorized out.
They are functions of the heavy quarks masses and of the scale parameter
 $\Lambda_{\rm QCD}$.
These QCD corrections are available at the
NLO~\cite{Buras:1990fn,Herrlich:1994yv,Herrlich:1995hh,Herrlich:1996vf} 
in the strong and
electromagnetic couplings.

The scale-dependent common factor of the short-distance corrections 
is given by
\beq
b ( \mu ) = \left[\alpha_s\left(\mu\right)\right]^{-2/9}
\left( 1 - J_3 \frac{\alpha_s\left(\mu\right)}{4 \pi} \right)
\, ,
\label{wc}
\eeq
where $J_3$ depends on the $\gamma_5$-scheme used in the regularization. The
NDR and HV scheme yield, respectively:
\beq
J_3^{\rm NDR} = -\frac{307}{162} \quad \quad \mbox{and} \quad \quad
 J_3^{\rm HV} =-\frac{91}{162} \, .
 \eeq

All the other numerical inputs can be taken as in the $\Delta S=1$ case.

\section{Chiral Perturbation Theory}

Quarks are the fundamental hadronic matter. However, the particles we observe
 are those built out of them: baryons and mesons. In the sector 
of the lowest
mass pseudoscalar mesons (the would-be Goldstone bosons: $\pi$,
$K$ and $\eta$), the interactions can be described in terms of an effective
theory, the chiral lagrangian, that includes only these states.
The chiral lagrangian and chiral perturbation 
theory~\cite{Weinberg:1979kz,Gasser:1985gg,Gasser:1984yg}
provide  a faithful representation of 
this sector of the standard model  after the
 quark and
gluon degrees of freedom have been integrated out.  The form of
this effective field theory and all its possible terms
are determined by $SU_L(3) \times SU_R(3)$
chiral invariance and Lorentz invariance.
The parts of the lagrangian
 which explicitly break chiral invariance are introduced in terms
of the quark mass matrix $\cal{M}$. 

The strong chiral lagrangian is completely fixed to the leading order in
momenta by symmetry requirements and the Goldstone boson's decay constant $f$:
\beq
{\cal L}_{\rm strong}^{(2)} = 
\frac{f^2}{4} \Tr \left( D_\mu \Sigma D^\mu \Sigma^{\dag} \right)
 +  \frac{f^2}{2}
B_0 \Tr \left( {\cal M} \Sigma^{\dag} +  \Sigma {\cal M}^{\dag} \right) \, ,
\label{L2strong}
\eeq
where ${\cal M} = \mbox{diag} [ m_u, m_d, m_s ]$ and $B_0$ is given by
$\langle \bar{q}_i q_j \rangle = - f^2 B_0 \delta_{ij}$, with
\beq
\langle \bar{q} q \rangle = - \frac{f^2 m_K^2}{m_s + m_d}
                          = - \frac{f^2 m_\pi^2}{m_u + m_d} \; ,
\label{qqPCAC}
\eeq
according to PCAC in the limit of $SU(3)$ flavor symmetry
($f_K=f_\pi\equiv f$).
The $SU_L(3) \times SU_R(3)$ field 
\beq
\Sigma \equiv \exp \left( \frac{2i}{f} \,\Pi (x)  \right)
\label{sigma}
\eeq
contains the pseudoscalar octet:
\beq
\Pi (x)  = \frac{1}{2} \sum_{a=1}^8 \lambda^a \pi^a (x) = \frac{1}{\sqrt{2}}
\left[ \begin{array}{ccc} \tilde{\pi}^0 & \pi^+ & K^+ \\
                          \pi^- & -\bar{\pi}^0 & K^0 \\
                           K^- & \bar{K}^0 & \tilde{\pi}^8 \end{array}
\right]  \, ,
\eeq
where
\beq
\tilde{\pi}^0 = \frac{1}{\sqrt{2}} \pi^0 + \frac{1}{\sqrt{6}} \eta_8\, , \qquad
\bar{\pi}^0 = \frac{1}{\sqrt{2}} \pi^0 -
\frac{1}{\sqrt{6}} \eta_8\, , \qquad \tilde{\pi}^8 = -
\frac{2}{\sqrt{6}} \eta_8 \, .
\eeq
The coupling
$f$ is, to lowest order,  identified with the  pion decay constant $f_\pi$
(and equal to $f_K$ before chiral loops are introduced); it defines a
characteristic scale
\beq
\Lambda_{\chi} \equiv 2 \pi \sqrt{6/N_c} \, f \, \simeq 0.8\: \mbox{GeV} \, ,
\eeq
typical of the vector meson masses induced by the spontaneous breaking
of chiral symmetry.
 When the
matrix $\Sigma$ is expanded in powers of $f^{-1}$, the zeroth
 order term is the 
free  Klein-Gordon lagrangian for the pseudoscalar particles.

Under the action
of the elements $V_R$ and $V_L$ of the chiral
group $SU_R(3) \times SU_L(3)$, $\Sigma$ transforms linearly:
\beq
\Sigma ' = V_R \Sigma V_L^{\dag} \; ,
\eeq
with the quark fields transforming as
\beq
q_L ' = V_L \, q_L \qquad \mbox{and} \qquad q_R ' = V_R \, q_R \; ,
\eeq
and accordingly for the conjugated fields.

Quark operators are represented  in this language
in terms of the effective 
field $\Sigma$ and its derivatives. For instance, at the leading order,
the quark currents are given by
\bea
\bar q^j_L\gamma_\mu q^i_L & \to & -i \frac{f^2}{2} 
\left( \Sigma^{\dag} D_\mu \Sigma \right)_{ij} \, , \\
\bar q^j_R\gamma_\mu q^i_R & \to & -i \frac{f^2}{2}  
\left( \Sigma D_\mu \Sigma^{\dag} \right)_{ij} \, ,
\eea
while the quark densities can be written at $O(p^2)$ as
\bea
 \bar{q}^j_L q^i_R   & \to &  - 2 B_0
\left[ \frac{f^2}{4} \Sigma + L_5 \ \Sigma
D_\mu \Sigma^{\dag} D^\mu \Sigma +
4 B_0 L_8 \ \Sigma {\cal M}^\dag\ \Sigma \right]_{ij} \, , \nnu \\
 \bar{q}^j_R q^i_L   & \to &  - 2 B_0
\left[ \frac{f^2}{4} \Sigma^\dag + L_5 \ \Sigma^\dag
D_\mu \Sigma D^\mu \Sigma^{\dag} +
4 B_0 L_8 \Sigma^{\dag} {\cal M} \ \Sigma^{\dag} \right]_{ij} \, , 
\label{qLqR}
\eea
where $L_{5,8}$ are coefficients which belong to the $O(p^4)$ chiral
lagrangian. 
To the next-to-leading order in the momenta, in addition to the leading order
chiral lagrangian (\ref{L2strong}),
 there are ten chiral terms and thereby
ten coefficients
$L_i$ to be 
determined~\cite{Gasser:1984yg,Gasser:1985gg} either 
experimentally or by means of some
model. As we shall see, some of them play an important role in the physics 
of \ee. As an example, we display the $L_5$ and $L_8$ terms in
${\cal L}_{\rm strong}^{(4)}$ which appear in \eq{qLqR} and
govern much of the penguin physics:
\beq
L_5 \, B_0 \, \Tr \left[ D_\mu \Sigma D^\mu \Sigma^{\dag}
 \left( {\cal M} \Sigma^{\dag} +  \Sigma {\cal M}^{\dag} \right) \right] \quad
\mbox{and} \quad
L_8 \, B_0 \, \Tr \left[ {\cal M}^{\dag} \Sigma  {\cal M}^{\dag} \Sigma +
 {\cal M} \Sigma^{\dag}  {\cal M} \Sigma^{\dag} \right] \, .
\label{L4strong}
\eeq

\subsection{The Weak Chiral Lagrangian}

We can write  the most general
expression for the $\Delta S = 1$ chiral lagrangian in accordance with the 
$SU(3)_L\times SU(3)_R$ symmetry, involving
unknown constants of order $G_F$.
This is done order by order in the chiral
 expansion. Typical terms to $O (p^2)$ are obtained by inserting 
appropriate combinations of Gell-Mann matrices into the strong lagrangian.
The corresponding chiral coefficients
must then be determined by means of some model or by comparison to
the experimental data.

We find it convenient to write the $\Delta S=1$ chiral lagrangian at $O(p^2)$
in terms of the following eight terms, of which seven are linearly
independent:
\bea
{\cal L}^{(2)}_{\Delta S = 1}  = && G_{LR}^{(0)}(Q_{7,8})
\Tr \left( \lambda^3_2 \Sigma^{\dag} \lambda^1_1 \Sigma
\right) \nnu \\
&+& G_{LR}^{(m)} (Q_{7,8})\ \left[ \Tr \left( \lambda^3_2 
\Sigma^\dag \lambda _1^1 \Sigma {\cal M}^\dag \Sigma \right)
+ \Tr \left(\lambda_1^1 \Sigma  \lambda^3_2  \Sigma^\dag 
{\cal M} \Sigma^\dag \right)\right]  \nnu \\
&+& G_{\underline{8}} (Q_{3-10}) \Tr \left( \lambda^3_2 D_\mu \Sigma^{\dag}
D^\mu \Sigma\right)   \nnu \\
&+&  G_{LL}^a (Q_{1,2,9,10}) \, 
\Tr \left(  \lambda^3_1 \Sigma^{\dag} D_\mu \Sigma \right)
\Tr \left( \lambda^1_2 \Sigma^{\dag} D^\mu  \Sigma \right) \nnu \\
&+& G_{LL}^b (Q_{1,2,9,10})\, \Tr \left( \lambda^3_2 \Sigma^{\dag} D_\mu
\Sigma \right)
\Tr \left(  \lambda^1_1 \Sigma^{\dag} D^\mu \Sigma \right) \nnu \\
&+&  G_{LR}^a (Q_{7,8})\, \Tr \left( \lambda^3_2 
D_\mu  \Sigma \lambda^1_1  D^\mu \Sigma^{\dag} \right)  \nnu \\
&+&  G_{LR}^b (Q_{7,8})\, \Tr \left( \lambda^3_2 \Sigma^{\dag} D_\mu
\Sigma \right)
\Tr \left(  \lambda^1_1 \Sigma D^\mu \Sigma^{\dag} \right)  \nnu \\
&+&  G_{LR}^c (Q_{7,8}) \left[ \Tr \left( \lambda^3_1 \Sigma \right)
\Tr \left( \lambda^1_2  D_\mu
\Sigma^{\dag} D^\mu \Sigma\ \Sigma^{\dag} \right) 
+ \Tr \left( \lambda^3_1  D_\mu \Sigma D^\mu \Sigma^{\dag}\ \Sigma\right)
\Tr \left(  \lambda^1_2 \Sigma^{\dag} \right)  \right]
 \label{Lchi} \, ,
\eea
where $\lambda^i_j$ are combinations of Gell-Mann
$SU(3)$ matrices defined by $(\lambda^i_j)_{lk} = \delta_{il}\delta_{jk}$
and $\Sigma$ is defined in \eq{sigma}.
The covariant
derivatives in \eq{Lchi} are taken with respect to the external
gauge fields whenever they are present. Other terms are possible, but they 
can be reduced to these by means of trace identities.

The  non-standard form and notation of \eq{Lchi} is chosen to remind us 
of the flavor and
chiral structure of the effective four-quark operators which are represented
by the various terms.
In particular, in
$G_{\underline{8}}$ we collect
the $(\underline{8}_L \times \underline{1}_R)$ part of the interaction
which is induced by the
gluonic penguins and by the analogous components of the electroweak operators
$Q_{7-10}$. The two terms
proportional to $G_{LL}^a$ and $G_{LL}^b$ are an admixture of
the $(\underline{27}_L \times \underline{1}_R)$ and the
$(\underline{8}_L \times \underline{1}_R)$ part of the interactions
induced by the left-handed
current-current operators $Q_{1,2,9,10}$. 
The term proportional to $G_{LR}^{(0)}$ is
the constant (non-derivative) $O(p^0)$ part arising from the isospin
 violating
 $(\underline{8}_L \times \underline{8}_R)$ electroweak operators.
The $O(p^2)$ corrections  to $G_{LR}^{(0)}$
 are the quark mass term proportional
to $G_{LR}^{(m)}$ (related to $L_8$), the momentum corrections proportional 
 to $G_{LR}^{c}$ (related to $L_5$) and $G_{LR}^{a,b}$. 
One may verify that $G_{LR}^{(m)}$ and $G_{LR}^{c}$ can be obtained
by multiplying the bosonized expression of a left- and a right-handed
quark density (in a manner similar to $Q_6)$, 
while $G_{LR}^{b}$ is obtained as the
product of a left- and a right-handed quark current. It is therefore
natural to call these terms factorizable (although $G_{LR}^{b}$ has
a non-factorizable contribution).
 The term $G_{LR}^{a}$ is, however, genuinely 
non-factorizable~\cite{Fabbrichesi:1996iz}.

The terms proportional to $G_{\underline{8}}$, $G^a_{LL}$ and $G^b_{LL}$ have
been studied in the literature~\cite{Cronin:1967jq,Pich:1991mw,Bijnens:1993uz,Ecker:1993de} in the framework of chiral
perturbation theory. The three terms are not independent.
Those proportional to $G^a_{LL}$ and $G^b_{LL}$
can be written in terms of the $\underline{8}$ and $\underline{27}$
$SU(3)_L$ components as follows:
\beq
 {\cal L}_{\underline{27}} = 
G_{\underline{27}}(Q_i) \: \left[ \frac{2}{3}\,  
\Tr \left( \lambda^3_1 \Sigma^{\dag} D^\mu \Sigma \right)
\Tr \left( \lambda^1_2 \Sigma^{\dag} D_\mu \Sigma \right)
  + \: \Tr \left( \lambda^3_2 \Sigma^{\dag} D_\mu \Sigma \right)
 \Tr \left(  \lambda^1_1 \Sigma^{\dag} D^\mu \Sigma \right) \right] \, ,
\label{a27}
\eeq
which transforms as $(\underline{27}_L \times \underline{1}_{R})$, and
\beq
 {\cal L}_{\underline{8}} = 
G_{\underline{8}}(Q_i) \: \left[ 
\Tr \left( \lambda^3_1 \Sigma^{\dag} D^\mu \Sigma \right)
\Tr \left( \lambda^1_2 \Sigma^{\dag} D_\mu \Sigma \right)
  - \: \Tr \left( \lambda^3_2 \Sigma^{\dag} D_\mu \Sigma \right)
\Tr \left(  \lambda^1_1 \Sigma^{\dag} D^\mu \Sigma \right) \right] \, ,
\label{a8}
\eeq
which transforms as $(\underline{8}_L \times \underline{1}_{R})$.
We prefer to keep the  $\Delta S = 1$ chiral Lagrangian in the form given in
\eq{Lchi}, which makes the bosonization of each quark operator more
transparent, and
perform the needed isospin projections at the level of the matrix elements.
Equations (\ref{a27})--(\ref{a8}) provide anyhow the
comparison to the standard notation. 
The chiral coefficients in the two bases
are related by
\bea
G_{\underline{8}} (Q_{i}) & = & \frac{1}{5} \left[ 3 \: G^a_{LL}(Q_{i}) -
2 \:
G^b_{LL} (Q_{i}) \right]  \nnu \\
G_{\underline{27}} (Q_{i}) & = & \frac{3}{5} \left[ G^a_{LL}(Q_{i}) +
G^b_{LL} (Q_{i}) \right] \; ,
\eea
for $i=1,2$.
Notice that there is 
no over-counting of the $\underline{8}_L\times \underline{1}_{R}$ 
contributions to \eq{Lchi}
from the operators $Q_{9,10}$ when
a consistent  prescription like that given in \cite{Antonelli:1996nv}
is followed.

Concerning the $(\underline{8}_L \times \underline{8}_R)$ part
of the $\Delta S=1$ chiral lagrangian, the constant 
term was 
first considered in~\cite{Bijnens:1984ye}, while its mass and $O(p^2)$
momentum corrections were first 
discussed in~\cite{Antonelli:1996nv,Bertolini:1997nf}.

As an example of the form of the chiral coefficients, 
we give the determination in the leading order in
$1/N_c$ of the two most important contributions to \ee:
\beq
 G_{\underline{8}} (Q_{6}) = - 24 
\frac{\langle \bar q q \rangle ^2 L_5}{f^2} \, C_6 
\eeq
and
\beq
G^{(0)}(Q_{8}) = - 3 \langle \bar q q \rangle ^2 \, C_8 \; ,
\eeq
where $C_{6,8}$ are the Wilson coefficients of the operators $Q_{6,8}$
at the matching scale $\mu$.

 The $\Delta S=1$ $O (p^4)$ Lagrangian is much more
complicated~
\cite{Kambor:1990tz,Esposito-Farese:1991yq,Ecker:1993de,Bijnens:1998mb} 
but we will not need its explicit form. In fact,
only certain combinations of coefficients from the $O(p^4)$ are required
in order to compute the relevant amplitudes  to this approximation.

The $\Delta S =2$ weak chiral lagrangian is simpler.
At the leading order
$O(p^2)$,  the $\Delta S=2$ weak chiral  lagrangian is given by only one term:
\beq
{\cal L}^{(2)}_{\Delta S = 2} = 
 G(Q_{S2})
 \Tr \left( \lambda^3_2  \Sigma  D_{\mu}\Sigma^{\dag} \right)
  \Tr \left( \lambda^3_2  \Sigma  D^{\mu}\Sigma^{\dag} \right)
\, . \label{ds2}
\eeq
The chiral coefficient is in this case given at the LO in $1/N_c$ by
\beq
G(Q_{S2}) = -\frac{f^4}{4} \, C_{2S} 
\, .
\label{cbs}
\eeq

\section{Hadronic Matrix Elements}

The estimate of the hadronic matrix elements must rely on 
 long-distance effects of QCD.
It is useful to encode the result of different estimates in terms of the
$B_i$ parameters that are defined in terms of the matrix elements
\beq
\langle Q_i \rangle _{0,2} \equiv \langle ( \pi \pi )_{(I=0,2)} | Q_i | K^0 
\rangle
\eeq
 as
\beq
B_i^{(0,2)} \equiv \frac{\Re \langle Q_i \rangle _{0,2}^{\rm model}}
{\langle Q_i \rangle _{0,2}^{\rm VSA}} \; , 
\label{Bi02}
\eeq
and
give the ratios between hadronic matrix elements in a model and those of the
vacuum saturation approximation (VSA). The latter is defined by 
factorizing the four-quark operators, inserting the vacuum state in all
possible manners (Fierzing of the operators included) and by
keeping the first non-vanishing term in the momentum expansion of each 
contribution.

 As a typical example, the matrix
element of $Q_6$ in the factorized version can be written as
the product of density matrix elements
\bea
\langle \pi^+ \pi^-|Q_6| K^0 \rangle & = &
 2 \, \langle \pi^+|\overline{u}\gamma_5 d|0 \rangle
\langle \pi^-|\overline{s} u |K^0 \rangle  -
2  \langle \pi^+ \pi^-|\overline{d} d|0 \rangle  \langle 0|\overline{s} 
\gamma_5 d |K^0 \rangle
\nnu \\
&& + 2 \, 
\left[ \langle 0|\overline{s} s|0 \rangle \, - \, 
\langle 0|\overline{d}d|0 \rangle
\right] \,   \langle \pi^+ \pi^-|\overline{s}\gamma_5 d |K^0 \rangle \, ,
\label{MQ6}
\eea
where the matrix elements like 
$\langle0|\,\overline{s} \gamma_5 u\, |K^+ \rangle$ and
$\langle \pi^+ |\,\overline{s} d\, |K^+ \rangle$ are obtained
from PCAC and the standard parameterization
of the corresponding currents,
$\langle 0|\,\overline{s} \gamma^\mu \left(1 - \gamma_5\right) u\,|K^+\rangle$ 
and 
$\langle \pi^+ |\,\overline{s} \gamma^\mu 
\left(1 - \gamma_5\right) u\,|K^+  \rangle$.
In the same way, the left-left currents operators can be written 
in the factorizable
approximation in terms of matrix elements of the currents.

Notice that the definition in \eq{Bi02} neglects the imaginary 
(absorptive) parts of the
hadronic matrix elements. Imaginary and real components, when multiplied by
the corresponding short-distance coefficients and summed over 
the contributing operators, should reproduce the global
phase of the amplitude arising from final state interactions. 
However, some approaches to hadronic matrix elements
do not account for absorptive contributions.
Therefore, in order to make the discussion of the $B_i$ factors of different 
models as homogeneous as possible, we 
propose the definition in \eq{Bi02}. Consistently with the use of such
a definition, extra overall $1/\cos\delta_{0,2}$
factors appear in the $I=0,2$ amplitudes, 
as discussed in Sect. VI.

\subsection{Preliminary Remarks}

The $B_i$ parameters depend in principle
on the
renormalization scale $\mu$ and therefore they should be given together
with the scale at which they are evaluated.

In this respect, in a truly consistent 
calculation of the hadronic matrix elements,
the cancellation of the unphysical renormalization scale 
and scheme dependence of the Wilson coefficients should 
formally be proven order by order in perturbation theory.

The only approach that 
fully satisfies these requirements is that based
on the lattice regularization (discussed in subsection F),
where the same theory, namely QCD, is used in both the short- and
the long-distance regimes and the matching only involves the 
different regularization schemes.

The M\"unchen phenomenological approach (discussed in subsection E) 
represents a clever attempt to address the
problem of a consistent calculation of \ee 
in a framework originally based on the $1/N_c$ expansion.
In this approach one extracts as much information as
possible on the hadronic matrix elements 
by fitting the $\Delta I = 1/2$ selection rule
at a fixed scale and in a given renormalization scheme.
The scale and renormalization scheme
stability of physical amplitudes 
can then be obtained using perturbation theory since
the matching scale between short- and long- distance calculations is
large enough ($\mu = m_c$) to lie inside the perturbative regime.
The phenomenological input allows for a direct determination
of the current-current matrix elements and indirectly of some of the
penguin matrix elements, thus reducing the number of free
parameters in the $\Delta S = 1$ effective lagrangian. 
On the other hand, the same fit does not give 
any information on the actual value (and scheme dependence)
of the $B_{6,8}$ parameters at the given scale, which are the most relevant
for determining \ee.

In the Trieste group approach (discussed in subsection G)
there is no attempt to prove formally
the consistency of the matching along the lines stated above.
The matching is done between QCD on the short-distance side and 
phenomenological models, the $\chi$QM and chiral perturbation theory,
on the long-distance side. In the long-distance calculation the scale
and renormalization scheme dependences appear naturally. It is then assumed 
that these unphysical dependences may satisfactorily
match those of the short-distance calculation. 
The fact that this assumption is numerically
verified (even beyond expectation), thus giving at the given order of the
calculation a stable set of predictions, and that it allows
for a complete calculation of all matrix elements
in terms of a few basic ``non-perturbative'' parameters, 
make this phenomenological analysis valuable.
The pattern of contributions which emerges and which leads to 
a satisfactory reproduction of the $\Delta I = 1/2$ rule
may be of help in other investigations. 
The major weakness of the approach is
the poor convergence of the chiral expansion at matching
scales of the order of the $\rho$ mass or higher, 
which are required by the reliability of the perturbative 
strong coupling expansion.

Very recently the Dortmund group (see subsection D)
has developed a systematic procedure
for matching short- and long-distance calculations, improving 
both technically and conceptually on the
original $1/N_c$ approach of \cite{Bardeen:1987uz}. On the other hand,
at the present status of the calculation, the scale stability
of the matching with the short-distance coefficients is for some
of the relevant observables ($\Delta I = 1/2$, $3/2$
amplitudes,  $\hat B_K$) quite poor \cite{Hambye:1997dh,Kohler:1997pg}.

\subsection{The Vacuum Saturation  Approximation}

According to the discussion above it is clear
that there is no theoretical underpinning for the consistency of the VSA; 
it is a convenient reference frame which is equivalent
 to retaining terms of
$O(1/N_c)$ in the $1/N_c$-expansion to the leading (non-vanishing)
 order in the momenta for all Fierzed forms of the operators.
Its application should in general not be pushed beyond leading order
in the strong coupling expansion. On the other hand, we find it useful
for illustrative purposes to use the VSA hadronic
matrix elements together with NLO Wilson coefficients in order to
exhibit some features of the long-distance calculation and
allow for a homogeneous comparison with the other estimates. 
For this purpose we will use in all
numerical estimates the Wilson
coefficients obtained in the HV scheme and set the matching scale 
at 1 GeV (see table \ref{numWcoefs}).

Some of the relevant VSA hadronic matrix elements depend on parameters
that are not precisely known. 
As a consequence, the
knowledge of the $B_i$ is not the whole story and, 
depending on 
assumptions, different
predictions of \ee may well differ even
starting from the same set of $B_i$. 
It is therefore important to define carefully
the VSA matrix elements. According to the standard bosonization
of currents and densities at $O(p^2)$ one obtains:
\bea
\langle Q_1 \rangle _0 & = & \frac{1}{3} X \left[ -1 + \frac{2}{N_c} \right]
\; , \\
\langle Q_1 \rangle _2 & = & \frac{\sqrt{2}}{3} X \left[ 1 + \frac{1}{N_c}
\right]  \; , \\
\langle Q_2 \rangle _0 & = & \frac{1}{3} X \left[ 2  - \frac{1}{N_c} \right]
\; , \\
\langle Q_2 \rangle _2 & = &  \frac{\sqrt{2}}{3} X \left[ 1 + \frac{1}{N_c}
\right] \; , \\
\langle Q_3 \rangle _0 & = & \frac{1}{N_c} X \; , \\
\langle Q_4 \rangle _0 & = & X \; , \\
\langle Q_5 \rangle _0 & = &  - \frac{16}{N_c}  \, 
\frac{\langle \bar{q}q \rangle ^2 L_5}{f^6} \, X \; , \\
\langle Q_6 \rangle _0 & = &  - 16 \, 
\frac{\langle \bar{q}q \rangle ^2 L_5}{f^6} \, X  \; , \\
\langle Q_{7} \rangle _{0} & = & \frac{2 \sqrt{3}}{N_c} \, 
\frac{\langle \bar{q} q \rangle ^2}{f^3} 
+ \frac{8}{N_c} \, 
\frac{\langle \bar{q}q \rangle^2 L_5}{f^6} \, X + \frac{1}{2} X \; , \\
 \langle Q_{7} \rangle _{2} & = & \frac{\sqrt{6}}{N_c} \,
 \frac{\langle \bar{q} q \rangle ^2}{f^3}
  - \frac{\sqrt{2}}{2} X \; , \\
\langle Q_8 \rangle _0 & = &  2 \, \sqrt{3} \, 
\frac{\langle \bar{q} q \rangle ^2}{f^3} + 8 
\frac{\langle \bar{q}q \rangle^2 L_5}{f^6} \, X   + \frac{1}{2 N_c}  X \; \\
\langle Q_8 \rangle _2 & = &
\sqrt{6} \, \frac{\langle \bar{q} q \rangle ^2}{f^3}
  - \frac{\sqrt{2}}{2 N_c} X \; , \\
\langle Q_9 \rangle _0 & = &  - \frac{1}{2} X \left[ 1 - \frac{1}{N_c}
 \right] \; , \\
\langle Q_9 \rangle _2 & = &   \frac{\sqrt{2}}{2} X \left[ 1 + \frac{1}{N_c}
\right] \; , \\
\langle Q_{10} \rangle _0 & = &   \frac{1}{2} X \left[ 1 - \frac{1}{N_c}
 \right] \; , \\
\langle Q_{10} \rangle _2 & = &   \frac{\sqrt{2}}{2} X \left[ 1 + \frac{1}{N_c}
\right] \; ,
 \eea
where
\beq
X \equiv \sqrt{3} f \left( m_K^2 - m_\pi^2 \right)\ . 
\eeq
In addition, from the $O(p^4)$
chiral lagrangian evaluation of $f_K/f_\pi$
one obtains, neglecting chiral loops,
\beq
L_5 = \frac{1}{4} \left( \frac{f_K - f_\pi}{f_\pi} \right) 
\frac{f^2}{m_K^2 - m_\pi^2} \, 
\label{L5ooN}
\eeq
while the quark condensate may be written 
in terms of the meson and quark masses using \eq{qqPCAC}.
The subleading $1/N_c$ terms arise from the Fierzing of the quark
operators via the $SU(N_c)$ relation (\ref{colorfierz}).

In a similar manner, in the case of the $\Delta S=2$ amplitude,
the scale-dependent $B_K$ parameter is defined by the matrix element 
\beq
\langle \bar{K^0} | Q_{S2} | K^0 \rangle = \frac{4}{3} f_K^2 m_K^2
B_K 
\, .
\label{bk}
\eeq              
The scale independent parameter $\hat B_K$ is defined by
\beq
\hat B_K = b(\mu) B_K(\mu) \, .
\eeq
In the VSA, for which $b(\mu) =1$, the value
\beq
\hat B_K = \frac{3}{4} \left[ 1 + \frac{1}{N_c} \right] 
\eeq
is found.

As it has been mentioned before, already at the level of the VSA, it
is necessary to know the value of $f$, \qq or, via PCAC, 
the value of quark masses. Specifically, unless otherwise stated,
we will assume as reference values for the input parameters in the VSA
$f=f_\pi$ and \qq$(1\ {\rm GeV})$ $= - (238\ \mbox{MeV})^3$, which corresponds
via \eq{qqPCAC} to $(m_u+m_d)(1\ {\rm GeV}) = 12$ MeV, or equivalently
to $(m_s+m_d)(1\ {\rm GeV}) = 157$ MeV.

Notice that the 
evaluation of the matrix elements of the operators $Q_{6-8}$ 
requires already at the VSA level the strong $O(p^4)$ chiral 
coefficient $L_5$.
For this reason, the determination of $B_6$
has been disputed in the
past~\cite{Dupont:1984mj,Gavela:1984ez,Donoghue:1984ba,Chivukula:1986du}.

We shall discuss  the numerical results of the $B_i$ factors in an improved
VSA model which includes
the complete $O(p^2)$ corrections to the leading momentum independent terms
in the $Q_{7,8}$ matrix elements. In the same model we will
show the effect of the inclusion of final state interactions.
Then, we will 
summarize the published results of the three most developed estimates: 
the M\"unchen phenomenological approach, the Roma 
numerical simulations on the lattice and, 
among possible effective quark models,
the chiral quark model (for which the complete set of operator
basis has been analyzed by the Trieste group).

The values quoted for the $B_i$ are taken at different scales so that they
cannot be directly
compared. Notice, however, the two most important parameters, 
namely $B_6$ and $B_8^{(2)}$
have been shown to depend  weakly on the renormalization
scale for $\mu\gtap 1$ GeV~\cite{Buras:1993dy}.

\subsection{A Toy Model: VSA$+$}

A comparison between the VSA matrix elements and the chiral 
lagrangian of \eq{Lchi} shows that none of the $O (p^2)$
terms proportional to $G^{(m)}$, $G^{a}_{LR}$ and $G^{c}_{LR}$
is included in the standard VSA. 
These contributions enter
as additional corrections to the $O (p^0)$ leading term in the  
matrix elements of the operators $Q_7$ and 
$Q_8$~\cite{Antonelli:1996nv,Bertolini:1997nf}. 
With the help of \eq{qLqR} and keeping all $p^2$ terms 
one obtains
\bea
\langle Q_{7} \rangle _{0} & = & \frac{2 \sqrt{3}}{N_c} \, 
\frac{\langle \bar{q} q \rangle ^2}{f^3} 
+ \frac{8}{N_c} \, 
\frac{\langle \bar{q}q \rangle^2 L_5}{f^6} \, X + \frac{1}{2} X 
+ \frac{16 \sqrt{3}}{N_c} \frac{\langle \bar{q} q \rangle ^2}{f^5} \left( 2 L_8
- L_5 \right) m_K^2\ ,  
\label{Q70} \\ 
 \langle Q_{7} \rangle _{2} & = & \frac{\sqrt{6}}{N_c} \,
 \frac{\langle \bar{q} q \rangle ^2}{f^3}
  - \frac{\sqrt{2}}{2} X
+ \frac{8 \sqrt{6}}{N_c} \frac{\langle \bar{q} q \rangle ^2}{f^5} \left( 2 L_8
- L_5 \right) m_K^2 \; , \\
\langle Q_8 \rangle _0 & =  & 2 \, \sqrt{3} \, 
\frac{\langle \bar{q} q \rangle ^2}{f^3} + 8 
\frac{\langle \bar{q}q \rangle^2 L_5}{f^6} \, X  
+ \frac{1}{2 N_c} X
+ 16 \sqrt{3} \frac{\langle \bar{q} q \rangle ^2}{f^5} \left( 2 L_8
- L_5 \right) m_K^2  \; , \\ 
\langle Q_8 \rangle _2 & = &
\sqrt{6} \, \frac{\langle \bar{q} q \rangle ^2}{f^3}
  - \frac{\sqrt{2}}{2 N_c} X 
+ 8 \sqrt{6} \frac{\langle \bar{q} q \rangle ^2}{f^5} \left( 2 L_8
- L_5 \right) m_K^2 \; ,
\label{Q82}
\eea
where we have neglected $m_\pi^2/m_K^2$ terms.
The
$O(p^2)$ wave-function renormalization has been included by multiplying
the $O(p^0)$ term by
\beq
\sqrt{Z_K} Z_\pi = 1 - 4\ L_5\ \frac{m_K^2 + 2 m_\pi^2}{f^2} \ .
\label{wf}
\eeq
In this toy model, which we call VSA$+$, we
neglect all chiral loop corrections, even though they are of
$O(p^2)$ on the constant 
term in the $\Delta S=1$ chiral lagrangian (all other chiral 
loop corrections are of $O(p^4)$). 
The parameter $f$ in the $O(p^0)$ terms
of \eqs{Q70}{Q82}
may be rewritten in terms of the renormalized
$f_K$ and/or $f_\pi$. At $O(p^2)$ such a rewriting is not unique.
For the purpose of the present discussion
we take, as in the standard VSA, $f=f_\pi$. 
The terms proportional to $2 L_8 - L_5$ represent the additional
corrections to the VSA matrix elements.

In order to obtain an
estimate of the combination $2 L_8 - L_5$ consistent with that of $L_5$
in eq. (\ref{L5ooN}), used in the VSA, we employ the 
mass relation~\cite{Gasser:1985gg}
\beq
\frac{m_K^2}{m_\pi^2} = \frac{m_s + \hat m}{2 \hat m}\left(1 + \Delta_M\right)
\; ,
\eeq
where $\hat m = (m_u+m_d)/2$ and, neglecting chiral loops,
\beq
\Delta_M = \frac{8}{f^2}\left(m_K^2-m_\pi^2\right)
           \left[2 L_8 - L_5\right]\ .
\eeq
Assuming PCAC to hold with degenerate quark condensates,
and keeping $f_K \neq f_\pi$, we then obtain 
\beq
2 L_8 - L_5 =  \frac{1}{8}\left[\frac{f_\pi^2}{f_K^2} - 1\right]
               \frac{f^2}{m_K^2-m_\pi^2} \; .
\eeq

The purpose of introducing the VSA$+$ model is to show the relevance
of the $O(p^2)$ corrections to the leading term for the $\vev{Q_8}_2$
 matrix element which is crucial in determining $\varepsilon'/\varepsilon$.
The coefficients $B_7$ and $B_8$ are modified from their VSA values
as shown in Table \ref{Bip2}. 
Their values are essentially independent on the value of \qq, because
of the smallness of the terms not proportional to the quark condensate. 

\vbox{
\begin{table}[thb]
\caption[]{The $B_i$ in the VSA$+$ model described
in the text. All other $B_i$ parameters are equal to unity.}
\begin{center}
\begin{tabular}{c c}
$B_7^{(0)} = B_8^{(0)}$ & $0.7$\\
$B_7^{(2)} = B_8^{(2)}$ & $0.6$\\
\end{tabular}
\label{Bip2}
\end{center}
\end{table}}

Much uncertainty in the present toy model
is hidden in the approximations made in giving $L_5$ and $L_8$. 
As an example,
a determination of these coefficients in chiral perturbation
theory including dimensionally regularized
chiral loops gives, at the scale $m_\rho$,
a $B_8^{(2)}$ greater than one~\cite{Fabbrichesi:1996iz}.

A discussion of the implications of the VSA$+$ model for \ee 
and a pedagogical comparison with the standard VSA
are presented in Sect. VI.

\subsection{$1/N_c$ Corrections}

Chiral-loop corrections are of order $1/N_c$
and of order $O(p^4)$ in the momenta (except for those of the
leading electroweak term that are of $O(p^2)$). They have been
included in the $1/N_c$ approach of \cite{Bardeen:1987uz} by means of
a cut-off regularization 
that is then matched to the short-distance renormalization scale 
between 0.6 and 1 GeV. 
The values thus
found ($B^{(0)}_1 = 5.2$,  $B^{(0)}_2 = 2.2$,  $B^{(2)}_1 = 0.55$) although
encouraging toward an explanation of the $\Delta I =1/2$ rule were 
still unsatisfactory in view of trusting the approach for
a reliable prediction of \ee. 

Along similar lines, the Dortmund group~\cite{Heinrich:1992en} 
included chiral corrections to the relevant operators $Q_6$ and
$Q_8$. They did not report explicit values for their $B_i$.
However, from their analysis it is clear that they find
 a rather large enhancement of $B_6$ and a suppression of $B_8$.
More recently \cite{Hambye:1998sm} 
have estimated these coefficients in a new study which pays special attention
to the matching between the renormalization scale dependence
of chiral loops, regularized by a cut-off, and the dimensionally
regularized Wilson coefficients. They find almost
no enhancement in the $B_6$ but a larger suppression of $B_8$. No new
calculation of \ee has appeared so far. Some of the relevant
observables, as $B_K$ and the $I = 0,2$ amplitudes,
show at the present status of the calculation
a quite poor scale stability \cite{Hambye:1997dh,Kohler:1997pg}, which may
frustrate any attempt to produce a reliable estimate of \ee.

The parameter $\hat B_K$ has been independently
estimated in the $1/N_c$ expansion with explicit cut-off 
by \cite{Bijnens:1995br}, finding values between 0.6 and 0.8. 

A systematic study of chiral-loop corrections in dimensional regularization
was performed first by~\cite{Kambor:1991ah} and more recently redone
using the $\overline{MS}$ scheme by the
Trieste group~\cite{Bertolini:1996tp,Bertolini:1997nf}.
The chiral-loop corrections also generate an absorptive part in the
amplitudes which should account for the
final state interactions. In any case, they seem to play an
important role in the determination of the hadronic matrix elements.

\subsection{Phenomenological Approach}

The phenomenological approach
 of the M\"unchen group~\cite{Buras:1993dy,Buras:1996dq} 
 writes all hadronic matrix elements in terms of just a handful of
$B_i$: $B^{(0)}_2$ for the $(V-A) \otimes (V-A)$ operators and
 $B_6$ and $B^{(2)}_8$ for  $(V-A) \otimes (V+A)$ operators.
This approach exploits in a clever manner 
the available experimental data on the amplitudes $A_0$ and $A_2$
in order to extract the (scheme dependent) values of $B^{(0,2)}_{1,2}$ 
and, via operatorial relations, of some of the penguin matrix elements,
while leaving
$B_6$ and $B^{(2)}_8$ as free input parameters to be varied within
given limits.

In particular, $B_{1,2}^{(2)}$ are obtained 
directly from the experimental value
\beq
\Re A_2 = 1.50 \times 10^{-8} \: \mbox{GeV} \; ,
\eeq
 via the matching condition at $\mu = m_c$ and the scale independence of 
the physical amplitude as
\beq
\langle Q_{1} \rangle _2 = \langle Q_{2} \rangle _2 = 
\frac{\Re A_2}{c \; z_+(m_c)} \; ,
\eeq
where $c = G_F V_{ud}V^*_{us}/\sqrt{2}$ and
$z_+$ is the real part of the Wilson coefficient of the operator $Q_1 + Q_2$; 
$B_{9,10}^{(2)}$ are then obtained
by using the operatorial
relation 
\beq
\langle Q_{9,10} \rangle _2 = \frac{3}{2} \langle Q_{1} \rangle _2 \; .
\eeq

$B_{1,4,9,10}^{(0)}$
are similarly expressed as functions of $B^{(0)}_2$ by means of other
operatorial relations and matching conditions at the charm-mass scale.
In fact, in the HV scheme at $m_c$ there are no penguin contributions 
to \CP conserving amplitudes and in the NDR the penguin
contamination is numerically small. Therefore one can write
\beq
\langle Q_1\rangle_0 = \frac{\Re A_0}{c\; z_1 (m_c)} -
\frac{z_2(m_c)}{z_1(m_c)}\langle Q_2\rangle_0 \, .
\eeq 
Finally, $B^{(0)}_2$ is also obtained under the plausible assumption
$\langle Q_2-Q_1\rangle \geq \langle Q_2+Q_1\rangle \geq 0$,
valid in all known non-perturbative approaches,
from the experimental value of 
\beq
\Re A_0 = 33.3 \times 10^{-8} \: \mbox{GeV} \, . 
\eeq

The following operatorial relations, which hold exactly in the HV scheme,
may then be used
\bea
\langle Q_4\rangle_0 & =& \langle Q_3\rangle_0 + \langle Q_2\rangle_0 -
 \langle Q_1\rangle_0 \; , 
\label{operel4} \\
\langle Q_9\rangle_0 & =& \frac{3}{2}\langle Q_1\rangle_0 -
\frac{1}{2} \langle Q_3\rangle_0  \; , 
\label{operel9} \\
\langle Q_{10}\rangle_0 & =& \langle Q_2\rangle_0 +
\frac{1}{2} \langle Q_1\rangle_0 - \frac{1}{2} \langle Q_3\rangle_0 \; .
\label{operel10}
\eea
It is important to recall that $B_3$ is taken equal to 1, 
which may be a rather crucial assumption in the determination
of $B_4$, as we shall see. 

After imposing that  $B_5 = B_6$ and $B^{(2)}_7 = B^{(2)}_8$,
this leaves us with only two free input parameters $B_6$ and
$B^{(2)}_8$ that are varied within 20\% from unity.

The parameter $B_K$ is pragmatically
taken to span from the central value of the lattice (see the next section)
to that of QCD sum rules~\cite{Narison:1995kd}. 

\vbox{
\begin{table}[thb]
\caption[]{The $B_i$ in the M\"unchen phenomenological approach.  
The results for $B_{1,2,9,10}$
are obtained by fitting the $\Delta I = 1/2$ selection
rule in $K\to \pi\pi$ decays at the matching scale $\mu = m_c$.
We show the values obtained in the HV scheme 
for the central value of $\Lambda_{\rm QCD}^{(4)}= 325$ MeV.
The value for $B_4$ is obtained by assuming $B_3 = 1$.
All the remnant $B_i$ are taken equal to 1 except for $B_6$ and 
$B_8^{(2)}$ that are varied within $\pm 20$\% from unity.
The parameter $\hat B_K$ is scale and renormalization scheme independent.}
\begin{center}
\begin{tabular}{c c}
$B^{(0)}_1$  & $13$ \\
$B^{(0)}_2$  &  $6.2$ \\
$B^{(2)}_1 =B^{(2)}_2 $  & $0.47$\\
$B_4$ & $5.2$\\ 
$B_9^{(0)}$ &$7.1$ \\
$B_{10}^{(0)}$ &$7.7$ \\
$B_9^{(2)}=B_{10}^{(2)}$ & $0.47$\\
$\hat B_K$ & $0.75 \pm 0.15$\\
\end{tabular}
\label{Biburas}
\end{center}
\end{table}}

\subsection{Lattice Approach}

The regularization of QCD on a lattice and its numerical simulation
is the most satisfactory theoretical approach to the computation of the
hadronic matrix elements 
(for a review see, for instance, \cite{Sharpe:1994xx}), and
should, in principle, lead to the most reliable estimates. However,
technical difficulties still plague this approach and
only some operators have been precisely determined
on the lattice. In addition, the use of approximations like quenching
make it very difficult to assess the effective uncertainty of the calculation.

Another problem of the approach is that it is still not possible
to directly compute  the $K \rightarrow \pi \pi$ amplitude in Euclidean space. 
It is therefore necessary to rely on chiral perturbation theory in order
to obtain the amplitude with two final pions from that with just one. 
In this sense even the lattice approach is not, 
at least for the time being, a first-principle procedure.
As a matter of fact,
when considering the complete $O(p^2)$ chiral lagrangian of \eq{Lchi}
a problem arises in so far as the term proportional to
$G^c_{LR}$ has a vanishing contribution to $K \rightarrow \pi$.

\vbox{
\begin{table}[thb]
\caption[]{The $B_i$ coefficients obtained in the Roma lattice calculation 
at the matching scale $\mu = 2$ GeV in the NDR scheme. The values of
$B_{1,2}^{(0,2)}$ are derived from the phenomenological fit of the
$\Delta I = 1/2$ rule. 
Accordingly, $B_4$ is varied in the range $1 \div 6$.  
All others $B_i$ are taken equal to 1.}
\begin{center}
\begin{tabular}{c c}
$B_{5,6}$& $1.0 \pm 0.2$\\
$B_7^{(2)}$ &$0.6 \pm 0.1$\\
$B_8^{(2)}$ & $0.8 \pm 0.15$\\
$B_9^{(2)}$ & $0.62 \pm 0.10$\\
$\hat B_K$ & $0.75 \pm 0.15$\\
\end{tabular}
\label{Bilattice}
\end{center}
\end{table}}

Table \ref{Bilattice} 
summarizes the values obtained by direct lattice computations 
and used by the Roma group~\cite{Ciuchini:1993tj,Ciuchini:1995cd}.
For the other
coefficients for which no lattice estimate is available, the
following ``educated guesses'' are used:
\begin{itemize}
\item $B_{3,7,8,9}^{(0)} = 1$,
\item $B_{4}$ in the range 1 to  6, in order to account for
the large values of $B_{1,2}^{(0)}$ needed to 
reproduce the $\Delta I = 1/2$ rule. 
\end{itemize}   

The parameter $B_K$ is consistently taken from the lattice 
estimates~\cite{Ciuchini:1995cd}. This determination
 gives in turn the value quoted in Table \ref{Bilattice}
for $B_9^{(2)}$ by means of the relation  $B_9^{(2)} = B_K$ which holds
if isospin-breaking corrections are neglected. 

Finally, because of the matching scale being at 2 GeV, also open charm
operators similar to $Q_{1,2}$ 
but with the strange quark replace by a charm quark ($Q_{1,2}^c$)
should be included and a value of $B^c_{1,2} = 0\div 0.15$ is
assumed. The \eqs{operel4}{operel10} are replaced by
\bea
\langle Q_4\rangle_0 & =& \langle Q_3\rangle_0 
+ \langle Q_2\rangle_0 - \langle Q_1\rangle_0 
+ \langle Q_2^c\rangle_0 - \langle Q_1^c\rangle_0 \; , 
\label{operelc4} \\
\langle Q_9\rangle_0 & =& \frac{3}{2}\langle Q_1\rangle_0 
- \frac{1}{2} \langle Q_3\rangle_0 + \frac{3}{2}\langle Q_1^c\rangle_0 \; , 
\label{operelc9} \\
\langle Q_{10}\rangle_0 & =& \langle Q_4\rangle_0 
+ \langle Q_9\rangle_0 - \langle Q_3\rangle_0 \; .
\label{operelc10}
\eea

The strength of the lattice approach is the direct evaluation of the
crucial matrix elements $\vev{Q_6}$ and $\vev{Q_8}_2$. On the other hand, 
while the lattice calculations of $B_8^{(2)}$ 
appear to have settled to reliable numbers,
there is still no solid prediction for 
$B_6$~\cite{Gupta:1998bm,Martinelli:1998hz},
and therefore the possibility of sizeable deviations from unity
remains open.

The values in table~\ref{Bilattice}, which are those used for the current
lattice estimate of \ee,
agree with more recent 
determinations~\cite{Kilcup:1997ye,Gupta:1997yt,Conti:1997qk} 
except for $\hat B_K$ for which the updated central values of 
0.92~\cite{Conti:1997qk} and
0.90~\cite{Sharpe:1997ih} are obtained.

\subsection{Chiral Quark Model}

Effective quark models of QCD can be derived in the framework of
the extended Nambu-Jona-Lasinio (ENJL) model of chiral symmetry
breaking~\citeaffixed{Bijnens:1996ww}{For a review, see, e.g.:}.
Among them is the chiral quark model ($\chi$QM)
\cite{Manohar:1984md,Espriu:1990ff}.
This model has a term 
\beq
 {\cal{L}}_{\chi \mbox{\scriptsize QM}} = - M \left( \overline{q}_R \; \Sigma
q_L +
\overline{q}_L \; \Sigma^{\dagger} q_R \right) \, ,
\label{M-lag}
\eeq
added to an effective low-energy QCD lagrangian whose dynamical degrees
of freedom are the $u,d,s$ quarks propagating in a soft gluon background.
The quantity $M$ is
interpreted as the constituent quark mass in mesons (current quark masses
are also included in the effective lagrangian).
The complete
operatorial basis in \eq{Q1-10} has been analyzed for $K\to\pi\pi$ decays,
inclusive of chiral loops and complete $O(p^4)$ 
corrections, by the Trieste
 group~\cite{Bertolini:1996tp,Bertolini:1997nf}. 

In the factorization approximation, the matrix elements of the four
 quark operators are written in terms of better known
quantities like quark currents and densities, as already shown in \eq{MQ6}.
Such matrix elements (building blocks) like the current matrix elements 
$\langle 0|\,\overline{s} \gamma^\mu \left(1 - \gamma_5\right) u\,|K^+(k)
\rangle$ and $\langle \pi^+(p_+)|\,\overline{s} \gamma^\mu
 \left(1 - \gamma_5\right) d\,|K^+(k)\rangle$ and the matrix elements 
of densities,
$\langle0|\,\overline{s} \gamma_5 u\, |K^+(k)\rangle$,
$\langle \pi^+(p_+)|\,\overline{s} d\, |K^+(k)\rangle$,  
 are  evaluated up to $O (p^4)$ within the model. 
The model dependence in
the color singlet current and density matrix elements appears 
(via the $M$ parameter) beyond the leading order in the momenta expansion,
while the leading contributions agree with the well known expressions in terms
of the meson decay constants and masses.

Non-factorizable contributions due to
soft gluonic corrections are included by 
using Fierz-transformations and
by calculating building block matrix elements
involving the color matrix $T^a$ (see \eq{colorfierz}):
\bea
\langle 0|\,\overline{s} \gamma^\mu T^a\left(1 - \gamma_5\right) u\,|K^+(k)
\rangle\ ,  \qquad \qquad 
\langle \pi^+(p_+)|\,\overline{s} \gamma^\mu
 T^a\left(1 - \gamma_5\right) d\,|K^+(k) \rangle \; .
\eea
Such matrix elements are non-zero for emission of gluons.
 In contrast to the color singlet matrix 
elements above, they are model dependent starting 
with the leading order.
Taking products of two such matrix elements and using the relation
\beq
 g_s^2  G^a_{\mu\nu}G^a_{\alpha\beta} =
\frac{\pi^2}{3}\langle \frac{\alpha_s}{\pi}GG \rangle
\left(\delta_{\mu\alpha}\delta_{\nu\beta} -
\delta_{\mu\beta}\delta_{\nu\alpha}\right)
\label{gluonaverage}
\eeq
makes it possible to express non-factorizable gluonic corrections
in terms of the gluonic vacuum condensate~\cite{Pich:1991mw}. 
The model thus
 parameterizes all amplitudes in terms of the quantities
  $M$,  \qq, and \GG .
Higher order gluon condensates are omitted.

The leading order (LO) ($O(p^0,p^2)$) matrix elements 
$\langle Q_i \rangle ^{LO}_{I}$ and
the next-to-leading order (NLO) ($O(p^2,p^4)$) corrections 
$\langle Q_i \rangle ^{NLO}_{I}$
 for isospin $I=0,2$ for the
 pions in the final state are obtained by
properly combining  the building blocks.
The total hadronic matrix elements up to $O (p^4)$ can then be 
written:
\beq
\langle Q_i(\mu) \rangle _{I} =
 Z_\pi \sqrt{Z_K}  
\left [\langle Q_i \rangle ^{LO}_{I} \, + \,  
\langle Q_i \rangle ^{NLO}_{I}(\mu) \right]  + a_i^I(\mu)\, ,
\label{hme}
\eeq
where $Q_i$ are the operators in \eq{Q1-10}, and $a_i^I(\mu)$
are the contributions from chiral loops (which include wave-function
renormalization). The scale dependence
of the $NLO$ terms comes from the perturbative running of the
quark masses.
The wave-function renormalizations $Z_K$ and $Z_\pi$ arise in the $\chi$QM
from direct calculation of the  $K \rightarrow K$ and 
$\pi \rightarrow \pi$ propagators.

The quantities
$a_i^I(\mu)$ represent the scale dependent meson-loop corrections 
which depend on the chiral quark model via the tree level chiral coefficients.
They have been included by the Trieste group
by consistently applying the $\overline{MS}$ scheme of
dimensional regularization.

At $O(p^2)$ the $Q_{5,6}$ and $Q_{7,8}$ matrix elements
contain the NLO coefficients $L_5$ and $L_8$, which within the 
chiral quark model are given by
\beq
L_5 = - \frac{f^4}{8 {\langle \bar{q} q \rangle }} \frac{1}{M} \left( 1 - 6
\frac{M^2}{\Lambda_\chi^2}\right) \, ,
\label{L5chiQM}
\eeq
and
\beq
L_8 =  - \frac{N_c}{16 \pi^2} \frac{1}{24} 
- \frac{f^4}{16 {\langle \bar{q} q \rangle } M} \left( 1 
+ \frac{M f^2}{{\langle \bar{q} q \rangle }}\right).
\label{L8chiQM}
\eeq

The hadronic matrix elements are matched with the NLO Wilson coefficients
at the scale $\Lambda_\chi \simeq 0.8$ and the scale dependence of the
amplitudes is gauged by varying $\mu$ between 0.8 and 1 GeV.
In this range
the scale dependence of \ee remains always below 15\%, thus giving
a stable prediction. The $\gamma_5$ scheme dependence, which arise from
the quark integration in the $\chi$QM is also found to numerically cancel
to a satisfactory degree that of the NLO Wilson coefficients, and the
predictions of \ee in the HV and NDR schemes differ only by 10\%. 
The results reported in the following are those of the HV scheme.

In order to restrict the possible values of the input parameters $M$,
\qq and \GG , the Trieste group has studied
 the $\Delta I = 1/2$ selection rule
for non-leptonic kaon decay within the $\chi$QM.  
By fitting at the scale $\mu = 0.8$ GeV the calculated amplitudes
to the experimental values, they find that within 
20\% error the $\Delta I = 1/2$ rule is reproduced for
\beq
M =  200 \; ^{+5}_{-3} \; \mbox{MeV} \, ,
\eeq 
\beq
 \langle \alpha_s GG/ \pi \rangle =  
\left( 334 \pm 4  \:\: \mbox{MeV} \right) ^4 \, ,
\label{range-gg}
\eeq
and 
\beq
\langle \bar q q \rangle = \left( -240^{+30}_{-10}\:\:\mbox{MeV} \right)^3\; .
\label{range-qq} 
\eeq
The fit is obtained for
values of the condensates which are in agreement with those
found in other approaches, i.e. QCD sum rules and lattice, although it is 
fair to say that the relation between the gluon condensate of
QCD sum rules and lattice and that of the $\chi$QM is far from obvious. 
The value of the constituent quark mass $M$ is in good agreement
with that found by fitting radiative kaon decays~\cite{Bijnens:1993xi}.

\vbox{
\begin{table}[thb]
\caption[]{The $B_i$ factors in the $\chi$QM. 
The results for $B_{1,...,10}$ are shown in the HV scheme,
at the scale $\mu = 0.8$ GeV,
for the central value of $\Lambda_{\rm QCD}^{(4)}= 340$ MeV.
The range in the matrix elements of the penguin operators $Q_{5-8}$
arises from the variation of \qq.
The value of the (scale and renormalization scheme independent) parameter
$\hat B_K$ includes the variation of all input parameters.}
\begin{center}
\begin{tabular}{c c}
$B^{(0)}_1$  & 9.5 \\
$B^{(0)}_2$  & 2.9 \\
$B^{(2)}_1 =B^{(2)}_2 $  & 0.41 \\
$B_3$ & $-2.3$ \\
$B_4$ & 1.9\\
$B_5 \simeq B_6$& $1.6 \pm 0.3$ \\
$B_7^{(0)} \simeq B_8^{(0)}$ &$2.5 \pm 0.1$ \\
$B_9^{(0)}$ &3.6 \\
$B_{10}^{(0)}$& 4.4 \\
$B_7^{(2)} \simeq B_8^{(2)}$ & $0.92 \pm 0.02$\\
$B_9^{(2)}=B_{10}^{(2)}$ &$0.41$ \\
$\hat B_K$ & $1.1 \pm 0.2$\\
\end{tabular}
\label{BiCQM}
\end{center}
\end{table}}

The obtained factors $B_{i}$ 
are given in Table \ref{BiCQM} in the HV
scheme, at $\mu= 0.8$ GeV, for the central value of 
$\Lambda_{\rm QCD}^{(4)}$~\cite{Bertolini:1997nf}.
The dependence 
on $\Lambda_{\rm QCD}$ enters, as for the M\"unchen approach,
indirectly via the fit of the
$\Delta I = 1/2$ selection rule and the determination of the parameters
of the model.
The uncertainty in the matrix elements of the penguin operators $Q_{5-8}$
arises from the variation of \qq. This affects sensibly the $B_{5,6}$
parameters because of the
linear dependence on \qq of the $Q_{5,6}$ matrix elements in the $\chi$QM,
contrasted to the quadratic dependence of the corresponding VSA
matrix elements. Accordingly, $B_{5,6}$ scale as 
$\langle \bar q q \rangle ^{-1}$, or via PCAC as $m_q$, and 
therefore are sensitive
to the value chosen for these parameters.
It should be remarked that
in the $\chi$QM analysis of~\cite{Bertolini:1997nf} the central value
of the quark condensate at the scale $\mu = 0.8$ GeV is given by 
$\langle \bar q q \rangle (0.8\ \mbox{GeV})= 
\left( -240\:\:\mbox{MeV} \right)^3$.
As a consequence, the VSA normalization, for the central value of the
quark condensate, numerically differs from that used in Sect. III.B, 
which corresponds to 
$\langle \bar q q \rangle (0.8\ \mbox{GeV})= 
\left( -222\:\:\mbox{MeV} \right)^3$. Finally,
it si interesting to notice that decreasing the value
of the quark condensate in the $\chi$QM depletes the 
$\langle Q_8 \rangle$ matrix element relatively to  $\langle Q_6 \rangle$,
and viceversa.  

The parameter $\hat B_K$ is scale and renormalization scheme independent
and the error given includes the variation of all input 
parameters~\cite{Bertolini:1997ir}.
 
Non-factorizable gluonic corrections are important
for the $CP$-conserving amplitudes (and account for the values of $B_1^{(0)}$
and $B_1^{(2)}$)  but are otherwise inessential in the determination of \ee.

\subsection{Discussion}

We would like to make a few comments on the  determinations of 
the matrix elements in the various approaches above.

\begin{itemize}

\item
All techniques attempt to take into account 
the $\Delta I =1/2$ rule, which is the most
preeminent feature of the physics of hadronic kaon decays. The  
direct fit of the rule 
in the phenomenological and lattice approaches determines
some of the hadronic matrix elements. In 
the $\chi$QM approach, the same fit
constrains the few input parameters of the model, 
in terms of which all matrix elements are expressed. The  $\chi$QM
approach is the only one for which the fit of the rule determines all 
hadronic matrix elements.

Since the operators $Q_1$ and $Q_2$, 
which dominate the  $\Delta I =1/2$ amplitude, 
do not enter directly the determination of \ee, 
the way such a fit affects \ee is
indirect and based on the use of operatorial relations as those
given in \eqs{operel4}{operelc10} 
in order to obtain information on the matrix elements of some
of the penguin operators.

According to \eq{operel4}
a large value of $ \langle Q_2 \rangle_0 - \langle Q_1 \rangle_0$ determines a 
proportionally large one for $ \langle Q_4 \rangle_0 $ 
if one assumes that $ \langle Q_3  \rangle_0$ has a positive
value. 
In the phenomenological approach 
$\langle Q_3\rangle_0 = 1$ is assumed thus
obtaining a rather large value for $B_4$. Similar values for $B_4$
are obtained, via a similar fit of the $\Delta I = 1/2$ selection
rule, in the lattice (see \eq{operelc4}). 
In the $\chi$QM, $B_3$ turns out to be  
large and negative and such that $B_4$ remains relatively small, albeit
larger than unity. At the same time the value of
$\langle Q_9 \rangle_0$ is increased.
The net effect is, by looking at the sign of the contributions of the
various operators depicted for the VSA in fig. \ref{pie}, an increase
of the predicted value of \ee.

It would be very interesting to have a lattice estimate of $B_3$ as a
 test of the two scenarios.

\item
The crucial parameters $B_6$ and $B_8^{(2)}$, 
are calculated in the lattice, in the $\chi$QM approach
at $O(p^4)$, and recently by a new estimate of the Dortmund group
in $1/N_c$ at $O(p^2)$ including chiral loops via a cut-off
regularization. 

The phenomenological approach
varies them according to a 20\% uncertainty around their VSA values.

The $\chi$QM finds a substantially larger value for $B_6$ compared to the
other approaches. This is due to the meson-loop
enhancement of the $A_0$ amplitude~\cite{Kambor:1990tz,Antonelli:1996nv}.
It is an open question
how much of this effect is accounted for in the quenched
approximation on the lattice. 
In addition, the lattice calculation of $B_6$ suffers from large 
renormalization uncertainties.

The Dortmund group originally found a large enhancement
for $B_6$ and suppression for $B_8^{(2)}$. In the latest and novel estimate
by \cite{Hambye:1998sm} they find almost no enhancement for $B_6$ and a 
strong suppression for $B_8^{(2)}$. One should wait
for a complete $O(p^4)$ calculation before drawing conclusions 
from the numerical comparison with the $\chi$QM results.

Both the phenomenological approach and the lattice do not 
include the $O(p^2)$ correction terms for the matrix elements
of the operators $Q_{7,8}$. 
The effect of these terms may be within the range
of the $B_{7,8}^{(2)}$ values these two approaches consider.
However, when these corrections are added, they may have the effect
of reducing $B_{8}^{(2)}$ thereby increasing the
central value of \ee.
All present calculations of $B_8^{(2)}$ agree on a value smaller than the
VSA result. 

\item
Those lattice computations
 that compute the $B_i$ from the $K \rightarrow \pi$ amplitude, and then
obtains the $K \rightarrow \pi \pi$ amplitude by means of the
 chiral lagrangian,
use an incomplete $O(p^2)$ lagrangian. In particular, the term
proportional to $G^c_{LR}$ has a vanishing contribution to the  
$K \rightarrow \pi$ amplitude, and in order to be determined, the knowledge
of   the  $K \rightarrow \pi \pi$ amplitude is required. 

\item
The parameter $\hat B_K$ is numerically the same in the phenomenological
 and lattice
approaches and smaller than the  $\chi$QM result. 
This parameter has always been a source
of disagreement among different estimates.
Recent lattice
determinations tend to assign a larger central value to $\hat B_K$,
closer to the VSA result ($\hat{B}_K \equiv 1$).

The different values of $\hat B_K$ used in the various approaches
lead, as we shall see, to different ranges for
the relevant combination of CKM matrix elements which enters the
determination of \ee (see section V).

\item The $\chi$QM model approach is the only one for which all matrix
elements are actually estimated---and up to the $O(p^4)$ in
the chiral expansion.  Of course this approach suffers from its
model dependence and the fact that the scale and renormalization scheme 
stability of the computed observables is a numerical feature
that is not formally proven 
(while the lattice and the M\"unchen phenomenological
estimates are in principle safe in this respect).  
On the other hand, it is the only approach in which the $\Delta I =1/2$ rule
is well reproduced  in terms of natural values of the
few input parameters when non-factorizable effects like
soft-gluon corrections  and meson-loops are included. 
These non-factorizable contributions are important 
in estimating \ee as shown by the relatively large value of $B_6$  and
in the interplay between the operators $Q_1$, $Q_2$, $Q_3$ and $Q_4$
(related by $Q_4 = Q_3 + Q_2 - Q_1$).   

\item Chiral-loop corrections give in general important contributions
to the hadronic matrix elements. 
A complete calculation of the hadronic matrix elements at $O(p^4)$ 
has been performed only in the framework of the  $\chi$QM so far. 
\end{itemize}

Of course, it is not sufficient to know the $B_i$ factors in order to predict
\ee, since the impact of the Wilson coefficients and other input
parameters must also be taken into 
account. As we shall see, the predictions depend 
crucially on the determination of the relevant CKM entries and the 
value assigned to $m_s$ (or, via \eq{qqPCAC}, 
the value of the quark condensate \qq ).

\section{The Relevant CKM Matrix Elements}

The ratio \ee, once the measured value of \eps is used,
turns out to be proportional to the combination of  CKM matrix
elements
\beq
\Im\ \lambda_t \equiv \Im V_{td}V^*_{ts} \; ,
\eeq
which, by using the Wolfenstein parameterization of \eq{KM},
can be written as
\beq
\Im\ \lambda_t \simeq \eta\ \lambda^5 A^2  = \eta\ |V_{us}| |V_{cb}|^2 \; ,
\label{imt}
\eeq
where $A = |V_{cb}|/\lambda^2$ and $\lambda = |V_{us}|$.

In order to restrict the allowed values of Im $\lambda_t$ we can solve
simultaneously  three equations. 

The first equation is derived from \eq{defeps} and
 gives the constraint from the experimental
value of \eps:
\beq
\eta \left( 1 - \frac{\lambda^2}{2} \right) \left\{ 
\left[1-\rho \left(1- \frac{\lambda^2}{2}\right) \right] 
 |V_{cb}|^2\ \eta_2 S(x_t) + \eta_3 S(x_x,x_t) - \eta_1 S(x_c) \right\}
\frac{|V_{cb}|^2}{\lambda^8} \hat B_K = \frac{|\varepsilon |}{C \;\lambda^{10}}
= 0.226 \; ,
\label{bound1}
\eeq
where
\beq
C = \frac{G_F^2 f_K^2 m_K^2 m_W^2}{3 \sqrt{2} \pi^2 \Delta M_{LS}} \; .
\eeq
In writing \eq{bound1} we have neglected in $\Im \lambda_c^*\lambda_t$ the
term proportional to
 $\Re \lambda_t/\Re \lambda_c$ which
is of $O(\lambda^4)$ and used the unitary relation $\Im \lambda_c^* =
\Im \lambda_t$.

Two more equations are
those relating $\eta$ and $\rho$ to measured entries
of the CKM matrix:
\bea
\eta^2 + \rho^2 & =&  \frac{1}{\lambda^2} \frac{|V_{ub}|^2}{|V_{cb}|^2} \; ,
\label{bound2}  \\
\eta^2 \left(1- \frac{\lambda^2}{2}\right)^2  + 
\left[1-\rho \left(1- \frac{\lambda^2}{2}\right) \right]^2
& = & \frac{1}{\lambda^2} \frac{|V_{td}|^2}{|V_{cb}|^2} \;  .
\label{bound3} 
\eea

The allowed values of $\eta$ and $\rho$ are thus obtained, given
\eps , $m_t$, $m_c$
and~\cite{Barnett:1996hr}  
\bea
 |V_{us}| & = & 0.2205 \pm 0.0018  \; , \\
 |V_{cb}| & = & 0.040 \pm 0.003  \; , \\
 |V_{ub}|/|V_{cb}| & = & 0.08 \pm 0.02 \; . 
\eea
For $|V_{td}|$ we can use the
bounds provided by the measured $\bar B_d^0$-$B_d^0$
mixing according to the relation~\cite{Buras:1997fb}
\beq
|V_{td}| = 8.8\cdot 10^{-3} 
\left[\frac{200\ {\rm MeV}}{\sqrt{B_{B_d}} F_{B_d}}\right]
\left[\frac{170\ {\rm GeV}}{m_t(m_t)}\right]^{0.76}
\left[\frac{\Delta M_{B_d}}{0.50/{\rm ps}}\right]^{0.5}
\sqrt{\frac{0.55}{\eta_B}}\ .
\label{Vtd}
\eeq

The theoretical uncertainty on the hadronic  
$\Delta S=2$ matrix element 
controls a large part of the uncertainty on the determination
of $\Im \lambda_t$.
 For the renormalization group invariant parameter
 $\hat B_K$  we  
take, as a reference for the following discussion,
 the VSA value with a conservative error of $\pm 30$\%.

The $\Delta S=2$ parameters $\eta_{1,2,3}$ obtained from QCD are known to the 
NLO~\cite{Buras:1990fn,Herrlich:1994yv,Herrlich:1995hh,Herrlich:1996vf}.
We compute them by taking
$\Lambda^{(4)}_{\rm QCD} = 340 \pm 40$ MeV, 
$m_b(m_b)=4.4$ GeV, $m_c(m_c) =1.4$ GeV and
 $m_t^{\rm (pole)} = 175 \pm 6$ GeV, which (in LO) corresponds
to $m_t(m_W) = 177 \pm 7$ GeV, where running masses
are given in the $\overline{MS}$ scheme. 
As an example, for central values of the parameters
we find at $\mu=m_c$
\beq
\eta_1 = 1.33\ ,\quad \eta_2 = 0.51\ ,\quad \eta_3= 0.44 \; .
\eeq

\begin{figure}
\epsfxsize=9cm
\centerline{\epsfbox{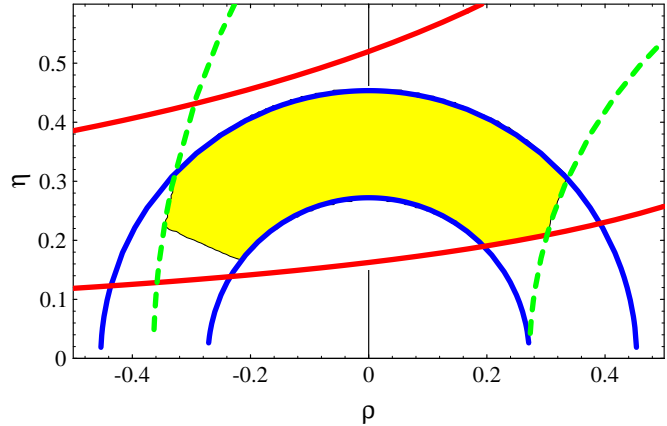}}
\caption{The allowed $\eta$ and $\rho$ ranges for 
         $\hat B_K = 1.0\pm 0.3$.}
\label{etarho}
\end{figure}

This procedure gives two possible ranges
for $\Im \lambda_t$
which correspond to having the CKM phase in
the I or II quadrant ($\rho$ positive or negative, respectively). 
Figure~\ref{etarho} gives the results of such an analysis for the
central value of $m_t$: the area enclosed
by the two black 
circumferences represents the constraint of \eq{bound2}, 
the area between the two gray (dashed)
circumferences is allowed by the bounds from
\eq{bound3}; the area enclosed by the two solid parabolic curves 
represents the solution of \eq{bound1} with
$\hat B_K$ in the $0.7\div 1.3$ range (notice that the upper
parabolic curve corresponds to the minimal value of $V_{cb}$ and
vice versa for the lower curve).

The gray region within the intersection of the curves is the range actually
allowed after the correlation in $V_{cb}$ between \eq{bound1} and
\eq{bound3} is taken into account. A further correlation is present in going
from $\eta$ to Im $\lambda_t$ in \eq{imt}. 

In the example of the VSA, where we have taken $\hat B_K = 1.0\pm 0.3$, 
from Fig. 1 we obtain
\beq
 0.51 \times 10^{-4} \leq \Im \lambda_t \leq 1.6 \times 10^{-4} \; .
\label{ImlamtVSA}
\eeq 

A further refinement of the analysis consists in assigning to each
pair of $(\rho,\ \eta)$ values a Gaussian weight according to the
deviations from the experimental central values of the computed
parameters $V_{ub}/V_{cb}$, $\Delta M_{B_d}$, $\varepsilon$. 
In this way, a Gaussian distribution of the uncertainty on $\Im \lambda_t$
(to be opposed to a flat one) is found and the error reduced.
We will use for the discussion of the VSA the flat result of \eq{ImlamtVSA}.

In general the renormalization group invariant parameter $\hat B_K$ depends on
the modeling of the hadronic matrix elements, so that different ranges of
$\Im \lambda_t$ should be expected according to the different approaches. 

\begin{itemize}

\item
In the M\"unchen phenomenological approach, where 
$\hat B_K=0.75 \pm 0.15$, a range
 \beq
 0.86 \times 10^{-4} \leq \Im \lambda_t \leq 1.71 \times 10^{-4}
 \label{imltphen}
 \eeq 
 is found for a flat-distribution of the uncertainties in the input
parameters, while the reduced range
 \beq
\Im \lambda_t = (1.29 \pm 0.22) \times 10^{-4} \label{im-ph}
 \label{imltphen2}
 \eeq
is obtained for a Gaussian treatment of the same uncertainties.
\item
In the Roma lattice calculation, which takes $\hat B_K=0.75 \pm 0.15$,
the range
\beq
\cos \delta_{CP} = 0.38 \pm 0.23 \ ,
 \label{imltlat}
\eeq
is obtained via the Gaussian treatment of the uncertainties,
where $\delta_{CP}$ is the CKM phase.
A result similar to \eq{im-ph} is found by means of 
\beq
\Im \lambda_t = |V_{cb}|^2 \frac{|V_{ub}|}{|V_{cb}|} 
\sqrt{ 1 - \cos ^2 \delta_{CP}} \; \, .
 \label{imltchiQM}
\eeq
\item 
In the Trieste $\chi$QM approach, which finds $\hat B_K=1.1 \pm 0.2$,
a flat scan of the input values leads to 
 \beq
 0.62 \times 10^{-4} \leq \Im \lambda_t \leq 1.4 \times 10^{-4} \; .
 \eeq
The larger value of $\hat B_K$ is responsible for the smaller values
$\Im \lambda_t$ obtained in this approach.
\end{itemize}
For a recent and detailed review on the determination of the CKM parameters
see \cite{Parodi:1998px}.

\section{Theoretical Predictions}

We have now all the ingredients necessary to understand the various
theoretical predictions for \ee . Let us first rewrite \eq{defeps'} in such a
way that the relationship with the effective
operators is more transparent.

The ratio \ee can be written as
\beq
 \frac{\varepsilon '}{\varepsilon} = 
e^{i \phi}\ 
\frac{G_F \omega}{2\mod{\epsilon}\Re{A_0}} \:
\mbox{Im}\, \lambda_t \: \:
 \left[ \Pi_0 - \frac{1}{\omega} \: \Pi_2 \right] \; ,
\label{epsprime2}
 \eeq
where, referring to the $\Delta S=1$ quark lagrangian of \eq{Lquark},
\bea
 \Pi_0 & = &  
\frac{1}{\cos\delta_0} \sum_i y_i \, \Re  \langle  Q_i  \rangle _0 
\ (1 - \Omega_{\eta +\eta'}) \; ,
\label{Pi_0}\\
 \Pi_2 & = & 
\frac{1}{\cos\delta_2} \sum_i y_i \, \Re \langle Q_i \rangle_2 \; .  
\label{Pi_2}\\
\eea

The phase of \ee is \cite{Maiani:1992ya}
\beq
\phi \, = \, \frac{\pi}{2} + \delta_0 - \delta_2 - \theta_\epsilon 
        = (0\pm 4)^0 \; ,
\label{phase}
\eeq
and we can take it as vanishing. We assume everywhere that $CPT$ is conserved.
An extra phase in addition to (\ref{phase}) would be present in the case
of $CPT$ non-conservation: present experimental bounds constrain it to be 
at most of the order of $10^{-4}$~(for a review see \cite{Maiani:1992ya}). 

Notice the explicit presence of the final-state-interaction phases
in eqs. (\ref{Pi_0}) and (\ref{Pi_2}). 
Their presence is a consequence of writing the absolute values 
of the amplitudes in term of their dispersive parts.
Theoretically, 
given that in \eq{Lqcoef} $\tau \ll 1$, we obtain
\beq
\tan\delta_I \simeq \frac{\sum_i z_i \, \Im \langle Q_i \rangle_I }
{\sum_i z_i \, \Re \langle Q_i \rangle_I}  \; .
\label{tandeltaI}
\eeq

A phenomenological estimate of the  rescattering phases can be
extracted from the elastic $\pi$-$\pi$ scattering.
In chiral perturbation theory to $O(p^4)$ one 
obtains~\cite{Gasser:1991ku}
\beq
\delta_0 - \delta_2 |_{s = m_K^2} = 45^0 \pm 6^0 \; .
\eeq
A more recent analysis of pion-nucleon collisions \cite{Chell:1993wu},
based on QCD sum rules and the extracted
s-wave $\pi-\pi$ isospin scattering lengths, finds at the 
kaon mass scale
\beq
\delta_0 = 34.2^0 \pm 2.2^0 \; , \quad \quad \quad 
\delta_2 = -6.9^0 \pm 0.2^0 \; ,
\eeq
and, accordingly,
\beq
\delta_0 - \delta_2 |_{s = m_K^2} = 41^0 \pm 4^0 \; .
\eeq
This result improves on older 
analyses~\cite{Basdevant:1974ru,Basdevant:1975aw,Froggatt:1977hu} for which
\beq
\delta_0 = 37^0 \pm 3^0 \; , \quad \quad \quad 
\delta_2 = -7^0 \pm 1^0 \; .
\eeq

All these results are consistent with each other and imply a
misalignment of the $I=0$ over the $I=2$ amplitude
by about 20\% ($\cos\delta_0/\cos\delta_2 \simeq 0.8$). 
Final state rescattering is not
included in the VSA hadronic matrix elements, and
in the lattice calculations, where the $K\to\pi$ 
transition is computed.
Absorptive components appear when
chiral loops are included, as in the $1/N_c$ approach of
\cite{Bardeen:1987uz} and in the $\chi$QM approach of the Trieste group.
In the latter framework the direct determination of the 
rescattering phases gives at $O(p^4)$ 
$\delta_0 \simeq 20^0$ and $\delta_2 \simeq -12^0$.
Although these results show features which are in qualitative
agreement with the phases extracted from pion-nucleon scattering, the 
deviation from the experimental data is sizeable, especially in the
$I=0$ component. On the other hand, at $O(p^4)$ the absorptive parts of the
amplitudes are determined only at $O(p^2)$ and disagreement with the
measured phases should be expected. 
At any rate, the effect of such
a discrepancy on \eqs{Pi_0}{Pi_2} is numerically 
reduced by the $\cos\delta_{0,2}$ dependence. 
The authors have therefore chosen to input
the experimental values of the rescattering phases
in all parts of their analysis. This amounts to overstimating systematically
the $I=0$ amplitude by about 15\%. Since the analysis
of the Trieste group is based on the fit of the $\Delta I = 1/2$
rule with a 20\% accuracy, such a bias is reabsorbed by the
uncertainty found in the determination of $\vev{\bar q q}$.

Since Im $ \lambda_u =0$ according to the standard conventions,
the short-distance component of $\varepsilon'/\varepsilon$
is determined by the Wilson coefficients $y_i$. 
Because, $y_1(\mu)=$ $y_2(\mu)=0$,
the matrix elements of $Q_{1,2}$ do not directly
enter the determination of $\varepsilon'/\varepsilon$.

We can take, as fixed input values:
\beq
\frac{G_F \omega}{2\mod{\epsilon}\Re{A_0}} \simeq 349 \ \mbox{GeV}^{-3} \; ,
\qquad 
\omega = 1/22.2 \; .
\label{values}
\eeq
The large value in \eq{values} for $1/\omega$ comes from
the $\Delta I =1/2$ selection rule. 

The quantity $\Omega_{\eta +\eta'}$, included in eq. (\ref{Pi_0})
for notational convenience, 
represents  the effect of the isospin-breaking 
mixing between $\pi^0$ and the etas, which generates a contribution
to $A_2$ proportional to $A_0$. 
$\Omega_{\eta + \eta'}$ can be written 
as~\cite{Donoghue:1986nm,Buras:1987wc} 
\beq
\Omega_{\eta + \eta'} = \frac{1}{3 \sqrt{2}} \frac{1}{\omega} \left[
\left( \cos \theta - \sqrt{2} \sin \theta \right)^2 +
\left( \sin \theta - \sqrt{2} \cos \theta \right)^2 
\frac{m_\eta^2 - m_\pi^2}{m_{\eta'} - m_\pi^2} \right] \frac{m_d - m_u}{m_s} 
\; ,
\eeq
where~\cite{Gasser:1985gg}
\beq
\frac{m_d - m_u}{m_s} = 0.022 \pm 0.002 \; .
\eeq
The mixing angle $\theta$ has been recently estimated
in a model-independent way~\cite{Venugopal:1998fq} to be
\beq
\theta = - 22^0 \pm 3.3^0 \; ,
\eeq
which is consistent with the values 
$\theta = -20^0 \pm 4^0$ found in chiral
perturbation theory~\cite{Donoghue:1986nm} 
and $\theta \simeq -22^0$ in the $1/N_c$ 
expansion~\cite{Gasser:1985gg}.

The values above yield
\beq
\Omega_{\eta + \eta'} = 0.28\ ^{+0.03}_{-0.04} \; .
\label{ometa}
\eeq
Smaller values are found once the uncertainty on the contribution of
the $\eta'$ is included~\cite{Cheng:1988dk}. 
For this reason, the more conservative range of values used in current
 estimates
of \ee is
\beq
\Omega_{\eta + \eta'} = 0.25 \pm 0.10 \ .
\label{ometaused}
\eeq

\subsection{Toy Models: VSA and VSA$+$}

Before summarizing the current estimates of \ee, it is useful to
study some of the steps through which
they are obtained in a toy model like that given by the VSA.
As already pointed out, this model, because of its simplicity,  
can be considered as a convenient reference framework against which
all other estimates are compared. 

The VSA$+$ model that 
we introduced in Section IV is an attempt to
improve on the VSA. It shows how
a more refined treatment of the electroweak operators,
which includes the $O(p^2)$ corrections to the leading constant term,  
can lead to a larger value of \ee. 

The main purpose of these toy models is
to illustrate in a simplified framework 
some general features of the calculation and
the impact of some assumptions on the predicted value of \ee.
As we have discussed in Section IV,
the VSA (as well as the VSA$+$) cannot give a reliable estimate because
of the absence of a consistent
scale and renormalization scheme matching with the NLO short-distance
QCD calculation.

In the present discussion we use the Wilson
coefficients in the HV scheme and set the 
reference value of the matching scale 
at 1 GeV (see table \ref{numWcoefs}).
We will then gauge the renormalization scheme
dependence of \ee
by varying the renormalization scheme from HV to NDR in the
VSA amplitudes.
Varying the matching scale around 1 GeV will show
the systematic uncertainty related to the choice of the renormalization
scale.

As we shall see,
different groups work at different renormalization scales because of the
peculiarities of their approaches. On the other hand,
in a consistent approach the
choice of the renormalization scale should be immaterial as long
as observables are concerned. The same holds for the scheme dependence.  

In addition to giving the $B_i$-parameters and the Wilson coefficients
in a common scheme and at a common scale, 
one needs to specify the numerical 
value for the input parameter \qq which appear in the penguin
matrix elements. 
We take the PCAC result, which at 1 GeV
and for $m_u + m_d = 12 \pm 2.5$ MeV gives
\beq
\vev{\bar q q} = (- 238\: ^{+19}_{-14} \: \mbox{MeV} )^3 \, .
\eeq
The mass  $m_s$ is often used instead of  $m_d$ and $m_u$. Such a change does
not reduce the error and may even add further uncertainties due to
the violations of PCAC that are larger in the $SU(3)$ case.

Each of the steps above, necessary in order to estimate \ee, may carry
in practice some model dependence and the reader must always
bear in mind the assumptions that have 
entered in the final numerical values.

Let us now study how the various operators come together to give the
final value of \ee. Figure~5 shows the individual contribution of each operator
in the VSA (gray histograms) and in the VSA$+$ (half-tone histograms).
The dark histograms show how the various contributions are affected
by changing the renormalization scheme from HV to NDR in
the VSA$+$ case.
 
The VSA and VSA$+$ estimates only differ in the $Q_{7,8}$ 
matrix elements, as already discussed in section IV.A,
while 
moving from HV to NDR affects mostly
the $Q_{5,6}$ contributions
(see Table \ref{numWcoefs}), thus  
leading to a potentially large effect
on the VSA prediction for \ee.

A central value of the order of $5 \times 10^{-4}$ is found in the 
VSA, whereas in going from VSA to VSA$+$ the
central value is increased by 50\%. 
A 25\% effect is then 
related to the renormalization scheme dependence in the VSA$+$,
which corresponds to a 50\% effect on the VSA.

\begin{figure}
\epsfxsize=9cm
\centerline{\epsfbox{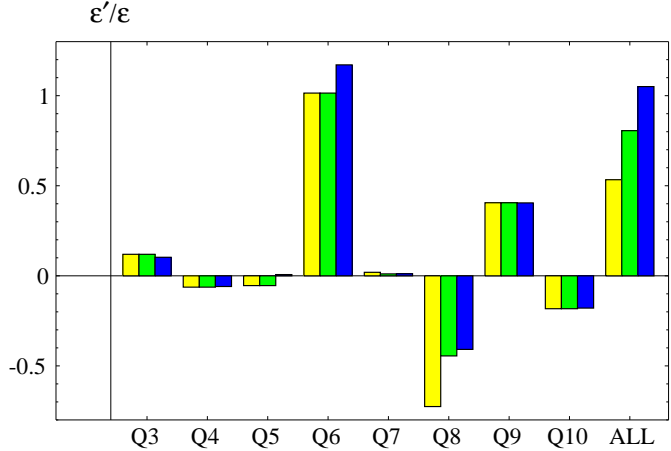}}
\caption{Anatomy of \ee in the VSA in units of $10^{-3}$ at $\mu =1$ GeV with
Im $\lambda_t = 1.1 \times 10^{-4}$. All other inputs are taken
at their central values. Depicted in gray is the VSA, in half-tone 
the VSA$+$ and in black the effect of changing the renormalization scheme
from HV to NDR.}
\label{charteps_n}
\end{figure}

Figure \ref{charteps_n} shows clearly how
systematic effects may sizeably move
the \ee value, due to the change in
the destructive interference between gluonic and electroweak penguins.

\begin{figure}
\epsfxsize=9cm
\centerline{\epsfbox{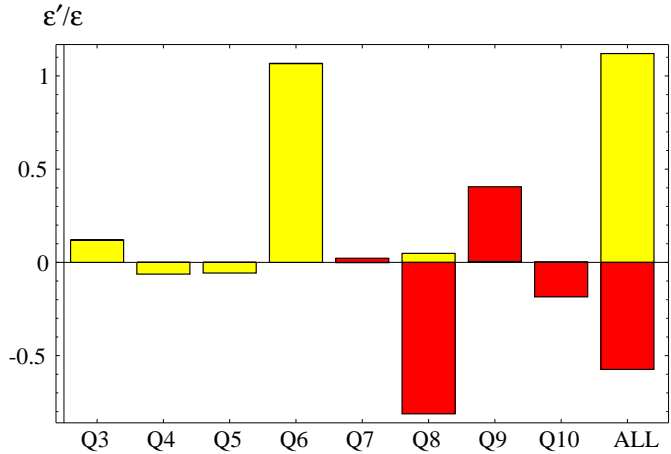}}
\caption{The distribution of the isospin $I=0$ (light gray) and $I=2$ 
(dark gray) contributions of each operator to \ee (in units of $10^{-3}$)
in the VSA.}
\label{charteps02}
\end{figure}

Figure \ref{charteps02} shows, for the case of the
VSA, the distribution of
the $I=0$ and 2 components in the contributions of each operator.
This figure is useful in disentangling the role and weight of the
individual operators according to the isospin projections.

\begin{figure}
\epsfxsize=12cm
\centerline{\epsfbox{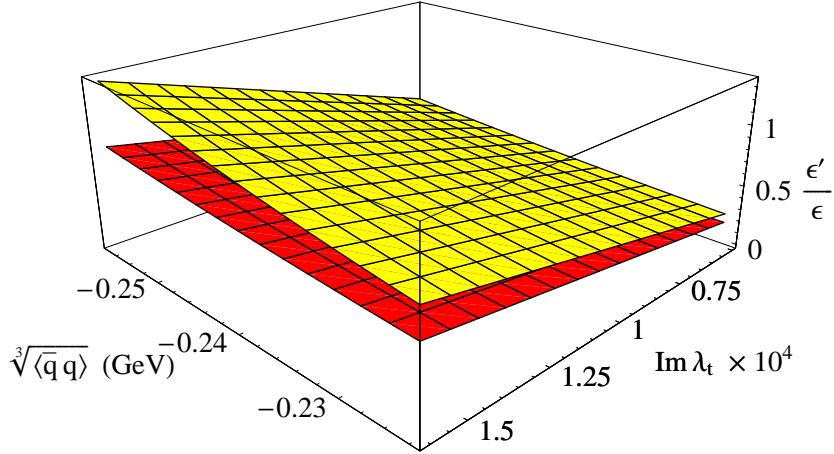}}
\caption{Parameter dependence of \ee in the VSA, in units of $10^{-3}$. 
The upper (lighter) surface corresponds to taking
$\mu =0.8$, $\Lambda_{\rm QCD} = 380$ MeV and $m_t = 161$ GeV, while 
the lower (darker) surface corresponds to  $\mu =1.2$ GeV,
 $\Lambda_{\rm QCD} = 300$ MeV and $m_t = 173$ GeV.}
\label{twoeps_vsa}
\end{figure}

Finally, in Fig.~\ref{twoeps_vsa}
the value of \ee in the VSA is shown as we continuously vary the two
most relevant parameters: Im $\lambda_t$ and \qq. The two surfaces
show in addition the
dependence of \ee on the short-distance input parameters
$\Lambda_{\rm QCD}$ and $m_t$ as we vary them between their $1\sigma$ limits, 
and include also the dependence on the matching scale which 
is varied from 0.8 to 1.2 GeV. 
Fig.~\ref{twoeps_vsa} is useful in showing the correlations
between the input parameters and \ee, which qualitatively
hold beyond the specific model considered. 

From Fig.~\ref{twoeps_vsa} we can finally extract the range 
of values taken by the parameter \ee in the VSA, in the HV scheme,
as we vary all relevant inputs. 
Taking into consideration the scale dependence of \qq
we find
\beq
\varepsilon'/\varepsilon = ( 0.5 \: ^{+0.6}_{-0.3} )
 \, \times 10^{-3} \; .
\label{VSA-HV}
\eeq 
Analogously, in the NDR scheme we obtain
\beq
\varepsilon'/\varepsilon = ( 0.8 \: ^{+1.3}_{-0.5} )
 \, \times 10^{-3} \; .
\label{VSA-NDR}
\eeq 
The large upper range of the NDR result is a consequence of the increase
of the scheme dependence of the Wilson coefficients as $\Lambda_{\rm QCD}$
increases and the renormalization scale decreases. 

While the toy models are useful in understanding how various possible
contributions enter in the final estimate of \ee, it is clear that some
important factors are not included. Among them, the actual range of
Im $\lambda_t$, strictly 
related to the determination of $\hat B_K$---which
might be quite different from the naive VSA---and 
the consistency of the
hadron matrix elements with the $\Delta I =1/2$ rule---which 
is important in assessing the confidence level of the 
\ee predictions. 
For this reason, we now turn to estimates that incorporate these important
features.

\subsection{Estimates of \ee}

There are three groups for which an up-to-date calculation is
available. In addition we will also briefly comment
on some recent partial results obtained within the $1/N_c$ approach.
 We will identify the various groups by the names
of the cities (M\"unchen, Roma and Trieste)
where most of the group members reside.

 In table~\ref{tab-inputs} we collect
some of the relevant inputs used by the three up-to-date estimates. 
There is overall agreement on the short-distance input parameters. 
The Trieste group differs from the other two in the value of $\hat B_K$,
and therefore of Im~$\lambda_t$, that is smaller, 
and for the inclusion of the FSI phases.
The matching scales are different because of the peculiarities of each 
approach which lead to the quoted energy scales. The scale 
(and renormalization scheme) dependence of 
the final estimates is however rather small.
We recall that while this stability is a formal property of the lattice
and M\"unchen phenomenological approaches, it is just a numerical
feature of the Trieste estimate.

\vbox{
\begin{table}[thb]
\caption[]{Comparison of input parameters in various approaches.
\label{tab-inputs}}
\begin{center}
\begin{tabular}{c c c c}
input & M\"unchen & Roma & Trieste \\
\hline
 $\Lambda_{\rm QCD}^{(4)}$ &  $325 \pm 80$ MeV &  $330 \pm 100$ MeV &  $340 \pm 40$ MeV \\
 $m_t(m_t)$ &  $167\pm 6$ GeV &  $167\pm 8$ GeV &  $167\pm 6$ GeV \\
 $m_b(m_b)$ &  4.4 GeV &  4.5 GeV &  4.4 GeV \\
 $m_c(m_c)$ &  1.3 GeV &  1.5 GeV &  1.4 GeV \\
 $\mu$ &  1.3 GeV &  2 GeV &  0.8 GeV \\ 
 $m_s(\mu)$ & $150\pm 20$ MeV &  $128\pm 18$ MeV & $220\pm 20$ MeV \\
 $\vev{\bar qq}(\mu)$ & via PCAC from $m_s$ & via PCAC from $m_s$ & 
        $(-240^{+30}_{-10}\ {\rm MeV})^3$ \\
 $\hat B_K$ &  $0.75 \pm 0.15$ &  $0.75 \pm 0.15$ &  
                                              $1.1 \pm 0.2$ \\
 $\Im \lambda_t \times 10^{4}$ &  $1.29 \pm 0.22$ &  $1.29 \pm 0.22$ &  
                                              $1.0 \pm 0.4$ \\
 $\cos\delta_0$ & 1 & 1 & 0.8 \\
 $\cos\delta_2$ & 1 & 1 & 1 \\
 $\Omega_{\eta+\eta'}$ &  $0.25 \pm 0.05$ & $0.25 \pm 0.10$ &  $0.25 \pm 0.10$ \\
\end{tabular}
\end{center}
\end{table}}

The experimental value of $m_t$ reported 
in Table~\ref{tab-inputs}---which has become available in the last few
years---greatly helps in
restricting the possible values of \ee and, as we shall see, rules out, at
least for a class of models, any mimicking of the superweak scenario by 
the standard model.

Starting with \eq{epsprime2}, and given the input
parameters in Table~\ref{tab-inputs},
the different estimates can be computed by writing \ee in terms of the VSA
to the matrix elements and the parameters $B_i$:
\bea
\sum_i y_i \vev{Q_i}_0 & = & X \left( y_4 B_4 + \frac{1}{N_c} y_3 B_3 \right)
- 16 \, \frac{\langle \bar{q}q \rangle ^2 L_5}{f^6} \, X \left( y_6 B_6 +
\frac{1}{N_c} y_5 B_5 \right)\nnu \\
& &  + \left( 2 \, \sqrt{3} \, 
\frac{\langle \bar{q} q \rangle ^2}{f^3} + 8 
\frac{\langle \bar{q}q \rangle^2 L_5}{f^6} \, X \right) \left( y_8 B_8 +
\frac{1}{N_c} y_7 B_7 \right) \nnu \\
 & &  + \frac{1}{2} X  \left( y_7 B_7 +
\frac{1}{N_c} y_8 B_8  \right)
-  \frac{1}{2} X \left[ 1 - \frac{1}{N_c} \right] \left( y_9 B_9 -
 y_{10} B_{10} \right) \; , 
\label{pi0} 
\eea
and
\bea
\sum_i y_i \vev{Q_i}_2 & = &   \sqrt{6} \,
 \frac{\langle \bar{q} q \rangle ^2}{f^3} \left( y_8 B_8 + \frac{1}{N_c} 
y_7 B_7 \right)
  - \frac{\sqrt{2}}{2} X \left(  y_7 B_7 + \frac{1}{N_c} 
y_8 B_8 \right)\nnu \\
& &  + \frac{\sqrt{2}}{2} X \left[ 1 + \frac{1}{N_c} \right] 
\left( y_9 B_9 + y_{10} B_{10} \right) \; .
\label{pi2}
\eea
In \eqs{pi0}{pi2} the values of the parameters $L_5$ and \qq
are obtained according to \eq{L5ooN} and \eq{qqPCAC} respectively,
taking into account the scale dependence of the quark masses.

By inserting the appropriated $B_i$, taking into account their
renormalization scheme dependence, 
the corresponding value of $\vev{\bar{q} q}$ (or $m_s$) and
the other short-distance inputs, varied within the given
uncertainties,
the reader can recover the estimates for the various 
groups that are reported in the next few subsections.

\subsubsection{Phenomenological Approach}

In the phenomenological approach of the 
M\"unchen group~\cite{Buras:1993dy,Buras:1996dq}
the matching scale is chosen at $\mu = m_c$ because it is
the scale at which penguins are decoupled from the \CP conserving
amplitudes and 
some of the $B_i$ parameters can be extracted from the knowledge of the
$\Delta I = 1/2$ rule.

In this approach all $B_i$ except 
$B_{3,5,6}$ and $B_8^{(2)}$
are determined from the experimental
values of physical processes. The
operator $Q_4$ receives an enhancement due to the rather large value used for
$B_4$  that comes from the fit of the $\Delta I =1/2$ rule with 
the assumption that $B_3=1$, as discussed in section IV.E.  

The parameters $B_6$
and $B_8^{(2)}$ are varied within a 20\% around the VSA values.
The quark condensate is written in terms of $m_s$, which 
is then varied according to the  uncertainty of its determination.

This procedure yields the two predictions~\cite{Buras:1996dq}
\beq
 - 1.2  \times 10^{-4} \leq \varepsilon '/\varepsilon
\leq 16.0 \times 10^{-4} \; ,
\label{Flat}
 \eeq 
for $m_s(m_c) = 150 \pm 20$ MeV and
\beq
 0  \leq \varepsilon '/\varepsilon
\leq 43.0 \times 10^{-4} \; ,
\label{lightmsFlat}
 \eeq 
for $m_s(m_c) = 100 \pm 20$ MeV. This second range is included
in order to study the implications of some recent lattice estimates 
of $m_s$ that found such a small values~\cite{Gupta:1997sa,Gough:1997kw}. 
Notice however
that the lower range is somewhat extreme in the light of more recent
lattice results now settling down at $m_s(\mbox{2 GeV}) = 110\pm 23$ 
MeV~\cite{Bhattacharya:1997ht}
(which corresponds to  $m_s(m_c) = 129\pm 27$ MeV). This range of
$m_s$ values is also consistent with recent QCD sum rules 
estimates~\cite{Colangelo:1997uy,Jamin:1997sa}.
On the other hand, a substantially larger determination of 
$m_s$ is obtained from the study of $\tau$ decays at LEP. A preliminary 
result from the ALEPH collaboration gives $m_s(m_\tau) = 172\pm 31$ 
MeV~\cite{Chen:1998mf}. It is therefore important for the determination
of \ee to understand better the value of this parameter, which 
via \eq{qqPCAC} affects
the size of the hadronic matrix elements of the most relevant operators.

For a Gaussian treatment of the uncertainties that affect the determination
of $\Im \lambda_t$, the values~\cite{Buras:1996dq}
\beq
 \varepsilon '/\varepsilon = (3.6 \pm 3.4) \times 10^{-4} \; ,
\label{Gauss}
\eeq
and
\beq
\varepsilon '/\varepsilon = (10.4 \pm 8.3) \times 10^{-4} \; ,
\label{lightmsGauss}
\eeq
are respectively found.

The same group also gives an approximated
analytical formula, in terms of the penguin-box expansion, that is useful in
discussing the impact in this estimate of the various input values:
\beq
\frac{\varepsilon '}{\varepsilon} = {\rm Im}\lambda_t \, F(x_t) \, ,
\label{eq:3}
\eeq
where
\begin{equation}
F(x_t) =
P_0 + P_X \, X_0(x_t) + P_Y \, Y_0(x_t) + P_Z \, Z_0(x_t) 
+ P_E \, E_0(x_t) \, .
\label{eq:3b}
\end{equation}

The $x_t$-dependent functions in (\ref{eq:3b}) are given, with
an accuracy of  better than 1\%, by
\begin{eqnarray}
X_0(x_t) = 0.660 \, x_t^{0.575}  \; , &\qquad&
Y_0(x_t) = 0.315 \, x_t^{0.78} \; , \label{eq:6} \\
Z_0(x_t) = 0.175 \, x_t^{0.93} \; , &\qquad&
E_0(x_t) = 0.564 \, x_t^{-0.51} \nonumber \; .
\end{eqnarray}

The coefficients $P_i$ are given in terms of $B_6^{(1/2)} \equiv
B_6^{(1/2)}(m_c)$, $B_8^{(3/2)} \equiv B_8^{(3/2)}(m_c)$ and $m_s(m_c)$
as follows
\begin{equation}
P_i = r_i^{(0)} + \left[ \frac{158 \: \mbox{MeV}}{m_s(m_c)+m_d(m_c)} \right]^2
\left(r_i^{(6)} B_6^{(1/2)} + r_i^{(8)} B_8^{(3/2)} \right) \, .
\label{eq:pbePi}
\end{equation}
The $P_i$ must be renormalization scale and scheme independent. They depend
however on $\Lambda_{\rm QCD}$. Table~\ref{tab:pbendr}, taken
from~\cite{Buras:1997fb}, gives the numerical
values of $r_i^{(0)}$, $r_i^{(6)}$ and $r_i^{(8)}$ for different values
of $\Lambda_{\rm QCD}^{(4)}$ at $\mu=m_c$.

\vbox{
\begin{table}[thb]
\caption[]{The penguin-box expansion coefficients 
for various $\Lambda_{\rm QCD}^{(4)}$ 
as given by~\cite{Buras:1997fb}. 
Only the coefficients $r_0$ depend at the NLO on
the renormalization scheme; the first raw gives their NDR values 
while the last row shows the corresponding values in the HV scheme. 
The results are given for $m_s(m_c) = 150$ MeV. 
\label{tab:pbendr}}
\begin{center}
\begin{tabular}{cccccccccc}
& \multicolumn{3}{c}{$\Lambda_{\rm QCD}^{(4)}=245$ MeV} &
  \multicolumn{3}{c}{$\Lambda_{\rm QCD}^{(4)}=325$ MeV} &
  \multicolumn{3}{c}{$\Lambda_{\rm QCD}^{(4)}=405$ MeV} \\
\hline
$i$ & $r_i^{(0)}$ & $r_i^{(6)}$ & $r_i^{(8)}$ &
      $r_i^{(0)}$ & $r_i^{(6)}$ & $r_i^{(8)}$ &
      $r_i^{(0)}$ & $r_i^{(6)}$ & $r_i^{(8)}$ \\
\hline
0 &
   --2.674 &   6.537 &   1.111 &
   --2.747 &   8.043 &   0.933 &
   --2.814 &   9.929 &   0.710 \\
$X$ &
    0.541 &   0.011 &       0 &
    0.517 &   0.015 &       0 &
    0.498 &   0.019 &       0 \\
$Y$ &
    0.408 &   0.049 &       0 &
    0.383 &   0.058 &       0 &
    0.361 &   0.068 &       0 \\
$Z$ &
    0.178 &  --0.009 &  --6.468 &
    0.244 &  --0.011 &  --7.402 &
    0.320 &  --0.013 &  --8.525 \\
$E$ &
    0.197 &  --0.790 &   0.278 &
    0.176 &  --0.917 &   0.335 &
    0.154 &  --1.063 &   0.402 \\
\hline
0 &
   --2.658 &   5.818 &   0.839 &
   --2.729 &   6.998 &   0.639 &
   --2.795 &   8.415 &   0.398 \\
\end{tabular}
\end{center}
\end{table}}

It is important to stress that the approximated formula~(\ref{eq:3b}), with
the numerical coefficient given in Table~\ref{tab:pbendr}, 
relies on the values of all $B_i$ used in the 
phenomenological approach, and great attention must be paid to the
possible effects of the different patterns of $B_i$ and the scale 
at which they are computed
when applying the same formula to other 
frameworks to compare predictions of \ee in the standard model.

\subsubsection{Lattice Approach}

In the lattice approach of the Roma 
group~\cite{Ciuchini:1993tj,Ciuchini:1995cd,Ciuchini:1997kd},
the matching scale is taken at $\mu = 2$ GeV.

As  it was for the M\"unchen group, the
operator $Q_4$ receives an enhancement due to the rather large value used for
$B_4$ in order to fit the $\Delta I=1/2$ rule with the assumption $B_3=1$.
The quark condensate is written in terms of $m_s$, which 
is then varied according to the  uncertainty of its determination. 

The parameters $B_6$
and $B_8^{(2)}$ are  explicitly  computed on the lattice,
although the determination of $\vev{Q_6}$ suffers from large
uncertainties (see section IV.D).

\begin{figure}
\epsfxsize=7cm
\centerline{\epsfbox{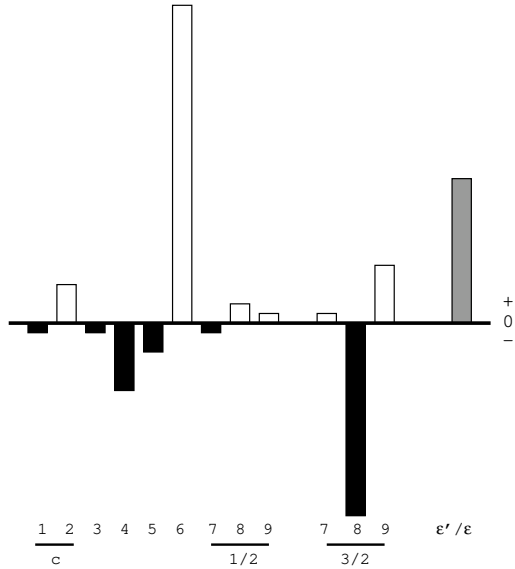}}
\caption{Anatomy of \ee in the lattice approach
in terms of the $I=0$ ($\Delta I = 1/2$) and $I=2$ ($\Delta I = 3/2$)
components.}
\label{ciuchini}
\end{figure}

Only the result obtained via the Gaussian treatment of the errors
in the input parameters is reported and yields~\cite{Ciuchini:1997kd}
\beq
\varepsilon '/\varepsilon = ( 4.6 \pm 3.0 \pm 0.4) \times 10^{-4} \, ,
\label{lat}
 \eeq 
where the first error is the variance of the distribution and the second
one comes from the residual $\gamma_5$-scheme dependence. Fig.~\ref{ciuchini}
from~\cite{Ciuchini:1997kd} 
shows the anatomy of \ee in the lattice case. In this figure, the various
contributions are shown in a manner similar to Fig.~5, with
the additional separation of the electroweak components
in isospin 0 and 2 amplitudes (as in Fig.~\ref{charteps02} for the VSA).

More recent estimates of $\hat B_K$ on the 
lattice~\cite{Kilcup:1997ye,Gupta:1997yt,Conti:1997qk}, find a value
larger than that used in deriving \eq{lat}, 
which makes Im $\lambda_t$ and, proportionally, \ee even smaller.

\subsubsection{Chiral Quark Model}

In the $\chi$QM approach of the 
Trieste group~\cite{Bertolini:1996tp,Bertolini:1997nf}, 
a rather low scale $\mu =0.8$ GeV is chosen 
because of the chiral-loop contribution that become perturbatively too large
at scales higher than $\Lambda_{\chi}\approx m_\rho$, 
the chiral-symmetry breaking scale. 
Such a low energy scale for the matching makes some of the Wilson coefficients
larger than in the other approaches and, correspondingly, more sensitive to
higher order corrections.

Let us also recall that the scale and renormalization scheme stability
of the computed observable is only a (welcomed) numerical feature and
no attempt to address formally the cancellation of unphysical
dependences is given.
On the other hand,  this estimate is the only one in which all $B_i$ are
computed within the same model and in terms of a few basic
parameters. It is also the only one for which
the full $O(p^4)$ amplitudes have been evaluated. It may therefore be
quite useful in complementing the other estimates by illustrating 
characteristic  patterns of the long-distance contributions. 

The value
of Im $\lambda_t$ is smaller than in the previous two estimates because of the
rather large value for $\hat B_K$ (see Table III) that is found in this model.

The quark condensate is a primitive input parameter that is varied
according to its determination in fitting the $\Delta I =1/2$ rule. The
value in \eq{range-qq} 
determined at the scale $\mu= 0.8$ GeV
by the Trieste group corresponds, via PCAC, 
to $m_s \simeq 220$ MeV. The quark masses appear explicitly
in the $\chi$QM calculation
at the NLO in momentum expansion and are treated
as independent parameters.
It is interesting to observe that in the $\chi$QM,
because of the
linear dependence on \qq of the $Q_{6}$ matrix element,
contrasted to the quadratic dependence of $\langle Q_8 \rangle$,
decreasing the value of the quark condensate depletes the destructive
interference between the two, and viceversa, partially compensating
for the overall change of scale.

Taking into account a 1-$\sigma$ flat distribution of
the input parameters the 
value~\cite{Bertolini:1997nf}
\beq
\label{TOnew}
\varepsilon '/\varepsilon = ( 1.7 \:^{+1.4}_{-1.0}) \times 10^{-3}
 \eeq 
is found.

\begin{figure}
\epsfxsize=9cm
\centerline{\epsfbox{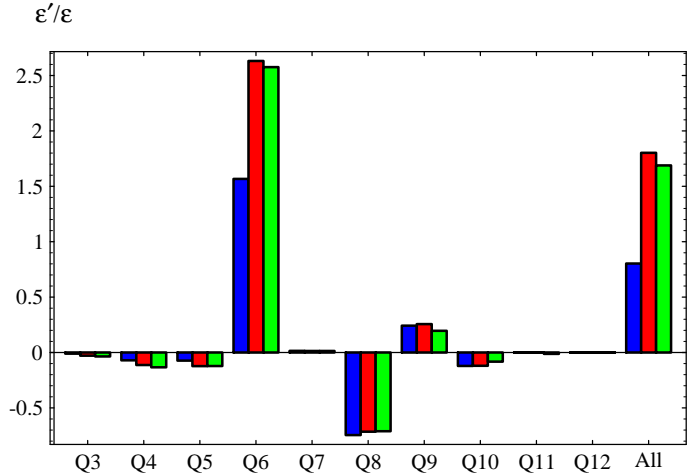}}
\caption{Anatomy of
\ee (in units of $10^{-3}$) within the $\chi$QM approach. 
In black the LO results (which includes the non-factorizable gluonic
corrections), in half-tone
the effect of the inclusion of chiral-loop corrections and in light gray
the complete $O(p^4)$ estimate.}
\label{charteps_qm}
\end{figure}

Figure~\ref{charteps_qm} shows explicitly the contributions of the
various operators, charted this time operator
by operator as in Fig~\ref{charteps_n}.

A previous estimate of \ee by the same group~\cite{Bertolini:1996tp} 
quoted the smaller value
\beq
 \varepsilon '/\varepsilon = ( 4 \pm 5 ) \times 10^{-4} \; .
\label{TOold}
\eeq
The change from (\ref{TOold}) to 
(\ref{TOnew}) is due to the following improvements: 
\begin{itemize}
\item Inclusion of the
      complete chiral lagrangian to $O(p^2)$ as discussed in section III.A; 
\item Extension of the matrix element calculation to the $O(p^4)$; 
\item Update of the short-distance analysis;
\item New ranges of input parameters as determined by the updated fit 
      of the $\Delta I =1/2$ rule~\cite{Bertolini:1997ir}.
\end{itemize}

\subsubsection{$1/N_c$ Approach}

The approach based on a $1/N_c$ estimate of the hadron matrix elements,
including chiral loops,
 has been first pursued by the M\"unchen 
group~\cite{Bardeen:1987uz,Buchalla:1990we}. 
Eventually, it was dropped in favor
of a phenomenological one that was judged to be better.

Successively, it was taken up by the 
Dortmund group~\cite{Paschos:1991as,Heinrich:1992en,Paschos:1996}.
Unfortunately, many details of  their work are not
available and there is no complete and updated calculation. 
For this reason we did not include it in Table~\ref{tab-inputs}.
 
The  latest available estimate quotes the value~\cite{Paschos:1996}
\beq
 \varepsilon '/\varepsilon = ( 9.9 \pm 4.1 ) \times 10^{-4} \, ,
\eeq
for $m_s(1\ {\rm GeV}) = 175$ MeV. 
This value is the result of a $B_6$ larger than 1
and a $B_8^{(2)}$ smaller than 1 as obtained by including chiral-loop
corrections in the matrix elements.

A very recent and new calculation of $B_6$ and $B_8$, 
which addresses systematically
the problem of a consistent 
renormalization scale matching between chiral loops and
Wilson coefficients,
yields a smaller value for $B_6$ while
a much suppressed value for $B_8^{(2)}$ is found~\cite{Hambye:1998sm}.
No new estimate of \ee has appeared yet.
However, some of the relevant
observables, as $B_K$ and the $I = 0,2$ amplitudes,
show at the present status of the calculation
a quite poor scale stability \cite{Hambye:1997dh,Kohler:1997pg}, which may
frustrate any attempt to produce a reliable estimate of \ee.

\subsection{\ee in the Standard Model: Summary and Outlook}

If we consider that  energy scales as different
 as $m_t$ and $m_\pi$ enter in an essential manner in the determination 
of the ratio \ee,
it is remarkable that this parameter can be predicted at all. 
Even more remarkable is the fact that all theoretical 
estimates are more or
less consistent and a well-defined window of possible values emerges.

\begin{figure}
\epsfxsize=10cm
\centerline{\epsfbox{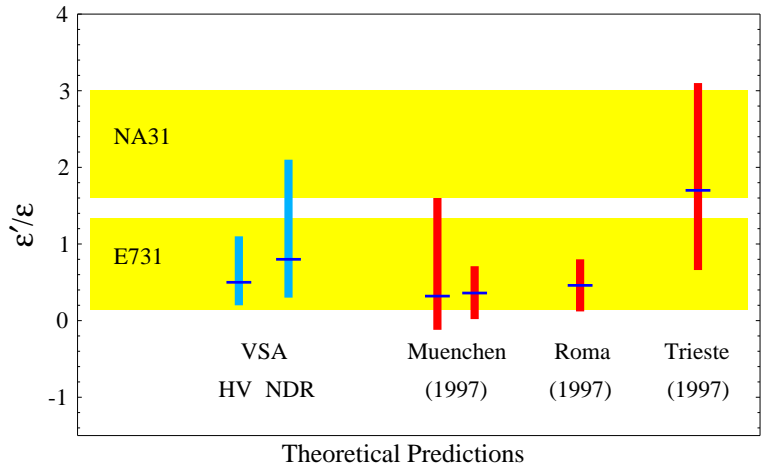}}
\caption{Current theoretical predictions for \ee in units of $10^{-3}$
in the standard model. The horizontal short bars 
mark the central values of each
prediction. The two gray areas correspond to the
current NA31 and E731 1-$\sigma$ experimental bounds. In dark gray the naive
VSA results are shown for comparison (the error bar 
includes a variation of the matching scale from 0.8 to 1.2 GeV).}
\label{fig}
\end{figure}

Figure~\ref{fig} collects the estimates we have discussed
and 
compares them with the
two present (1998) experimental ranges from CERN (NA31) and
FNAL (E731). 
We have also shown as a reference the results obtained in
the simple VSA, in the HV and NDR schemes, 
as discussed at the beginning of this Section.
We recall that the VSA error bars include a variation
of the matching scale from 0.8 to 1.2 GeV.

The two error bars depicted for the M\"unchen estimates correspond, from left 
to right, to flat and Gaussian scanning of the input data respectively.
Also the reduced size of the error bar of the lattice result
is due to the Gaussian treatment of the data.  

The entire range between zero and, roughly
$3 \times 10^{-3}$ is spanned by the available standard model predictions, 
thus dispelling the belief (that has been around in the last few years) 
that values of the order of $10^{-3}$ were difficult to 
account for within the standard model. 

Given the present theoretical and experimental results it is difficult 
to draw definite conclusions from their comparison beyond the fact
that there are no inconsistencies. On the other hand the forthcoming
experimental data may crucially help theorists in better
understanding the role of
non-perturbative QCD in the present estimates.
  
\begin{figure}
\epsfxsize=10cm
\centerline{\epsfbox{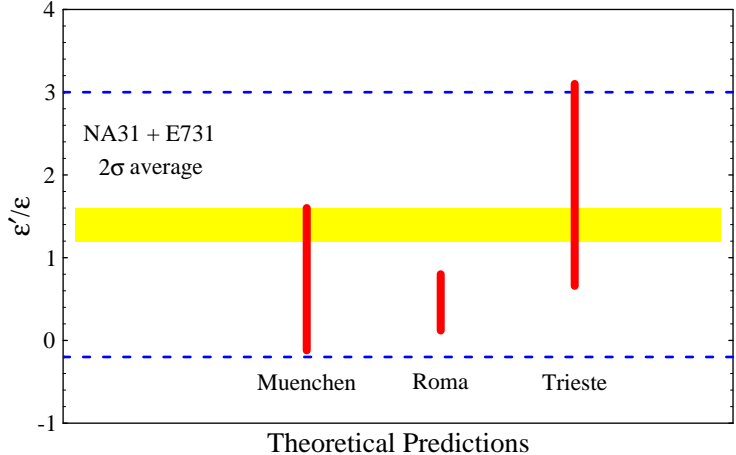}}
\caption{The combined
NA31 and E731 experimental bounds (area within the dashed lines),
are compared with the most updated
theoretical estimates for \ee (in units of $10^{-3}$).
The gray horizontal band represents
the future experimental sensitivity shown around the present 
experimental average value in \eq{average}.}
\label{fig2}
\end{figure}

To have a pictorial impression
of the dramatic improvement expected from the
currently running experiments one must shrink the
experimental ranges within 
a $\pm \: 2 \times 10^{-4}$ error band corresponding to two ticks on the
vertical scale of Fig.~\ref{fig}. 
This is shown in Fig.~\ref{fig2} by the horizontal gray band drawn on the
central value of the
$2\sigma$ average of the NA31 and E731 results
\beq
\varepsilon '/\varepsilon = (1.4 \ \pm 1.6) \times 10^{-3}\ ,
\label{average}
\eeq 
which is obtained by following the PDG procedure 
for error inflation when central values are in
disagreement~\cite{Winstein:1993sx}.

Such an improvement in the
experimental results will certainly spur a new wave of
theoretical analyses. We foresee at least three
directions along which such a re-analysis may take place:
\begin{itemize}
\item
Should the experimental results converge to a common error range
of the order of few $10^{-4}$, 
it will be useful to focus the attention on
the central values obtained by the various theoretical approaches in order
to better understand the most relevant effects at work.  
As an example, consider the case that the new experimental central value
turns out to be near or larger than the present average result.   
The comparison between the VSA and the VSA$+$ toy estimates discussed
in the present review, together with
the results of the Trieste group, suggest that the cancellation between
the gluon and electroweak penguin operators may be substantially reduced once 
$(i)$ the complete set of the electroweak $O(p^2)$ terms, and 
$(ii)$ higher order chiral corrections are taken into account. 
These effects can in part be included both in the M\"unchen 
estimate, for those matrix elements that are not determined 
phenomenologically, and in the lattice prediction.
Such effects may in fact  account for 
larger central values than those presently obtained in those estimates.
\item
For what concerns the reduction of the theoretical error, in all
estimates a crucial role is played by the knowledge of
the relevant CKM entries.
A large fraction of the theoretical error on \ee is related
to the uncertainty on $\Im \lambda_t$, which amounts
by inspection of \eqs{imltphen}{imltchiQM}, to a 30\%$-$40\% effect
on the total error.
The uncertainty on $\Im \lambda_t$ is presently dominated
by the  determination of $\hat B_K$. 
In this respect, a precise
determination of $\Im \lambda_t$ from $B$-physics alone---as expected 
from the upcoming $B$-factories and hadronic facilities---will 
free this part of the analysis from large hadronic uncertainties and thus 
reduce the impact of non-perturbative QCD in the
theoretical determination of \ee.
\item
Progress in the lattice estimate of hadronic elements is to be expected in
the next few years~\cite{Gupta:1998bm,Sharpe:1998hh}. 
A reliable estimate of the parameter $B_6$ is particularly
needed. 
In addition, achieving the needed  precision of the order
of 10\% or below in all relevant matrix elements implies
going beyond the quenching approximation. The inclusion of
higher-order chiral corrections can be also important.
Much work is beeing done at present
which indicates the possibility of addressing quantitatively this issue
in the near future. It is from lattice 
QCD that we should expect a conclusive word on the matter.  
\end{itemize}

\section{New Physics and \ee}

Physics beyond the standard model may enter
the determination of \ee in many ways. In particular, since
the origin \CP violation is still unclear,
\begin{itemize}
\item
It remains an open issue whether
the \CP violation observed in the $\bar K^0$-$K^0$ system
stems from complex Yukawa couplings or from a superweak
interaction which goes beyond the standard model; 
\item
Even maintaining that the observed \CP violation is not
superweak in nature, 
other sources of \CP violations may be present in addition to, or replace
the standard CKM phase in extensions of the standard model.
\item
Even if we insist that the CKM phase is the only source
of \CP violation, new particle contributions to the Wilson
coefficients of the effective quark Lagrangians still may
be relevant for the detailed prediction of \CP violating observables. 
\end{itemize}

However, given the discussion of the previous sections and considering
in particular the comparison between the present theoretical
and experimental results shown in Fig.~\ref{fig},
it appears to be a difficult task to disentangle new physics
effects in \ee.

Yet, one question that may be asked is whether
the present experimental window allows for visible signals
of non-standard physics.
In order to answer this question we may take the average
$2\sigma$ result of the NA31 and E731 experiments  
shown in Fig. \ref{fig2}
and compare it with the range obtained by the union
of the most recent theoretical estimates
(which is a reasonable, albeit biased, procedure).

It is clear that the case for observable signals of new physics is  marginal, 
to say the least, and
that, in order for new effects to become visible in \ee, 
the next run of experimental data must converge 
to the most unlikely areas of the present range,
thus pointing to values of \ee larger than 
a few times $10^{-3}$, thereby confirming the 2$\sigma$ upper range of
the NA31 result, or negative values, thus moving in the
lower side of the 2$\sigma$ E731 range.

For this reason, we think that it is not necessary to present an exhaustive
(and exhausting) review of all attempts to non-standard physics effects 
in \ee. 
The interested reader may 
consult~\cite{Grimus:1988kn,Winstein:1993sx,Nir:1997nj,Fleischer:1997km} 
for reviews of possible new-physics effects in \CP violation. 

It is nonetheless interesting to analyze
whether specific models affect the standard model prediction
via definite patterns.
In order to do so,
let us try to infer, inasmuch as possible in
a model independent way, 
how new physics may affect the
standard model prediction. 

\subsection{Model Independent Analysis}

The key ingredients for a theoretical prediction of \ee are
the determination of $\Im \lambda_t$, from the experimental
value of $\varepsilon$ and $B$-physics, and the calculation of all
 direct contributions to $\varepsilon'$. 
These depend, on the short-distance
side, from the values of the various components
of the Wilson coefficients and, on the long-distance side, 
on the value of $\hat B_K$ and the $\Delta S=1$ matrix elements
for $K\to\pi\pi$.

If we consider that the new effects modify
only the short-distance aspects of the analysis,
then the study of  $\varepsilon$ exhibits a general feature:
the new range of values for $\Im \lambda_t$ obtained
is always bounded from above by the maximum value given 
in Fig. 4 at $\rho=0$ by the $V_{ub}/V_{cb}$ measurement,
which is a tree level bound and therefore
robust to new effects.

As a consequence 
\begin{itemize}
\item
no sizeable enhancement of \ee with respect to the 
standard model estimate can be expected from a modification of the
short-distance part of $\varepsilon$. 
\end{itemize}
On the other hand,
the range of allowed values for $\Im \lambda_t$ may be substantially
reduced by new physics contributions, thus improving on the precision
of the \ee prediction.  

Acting on the matchings of the $\Delta S=1$ Wilson
coefficients $C_i$ in \eq{Lquark} at $\mu=m_W$ affects the final outcome
on \ee. There are patterns on how changing the $C_i(m_W)$
may affect the $y_i$ at the low energy scale ($\mu\simeq 1$ GeV)
via strong and electromagnetic renormalizations.

In Table \ref{Wcoefs}, we have schematically reported the distribution
of the different types of diagrams that 
determine the initial matching of the Wilson coefficients.
Since new heavy particles may show their presence through
their virtual exchange in the diagrams depicted in Fig. 1,
and different type of diagrams show different short distance
properties, it is important to keep an eye on how the 
relevant Wilson coefficients are generated

\begin{figure}
\epsfxsize=8cm
\centerline{\epsfbox{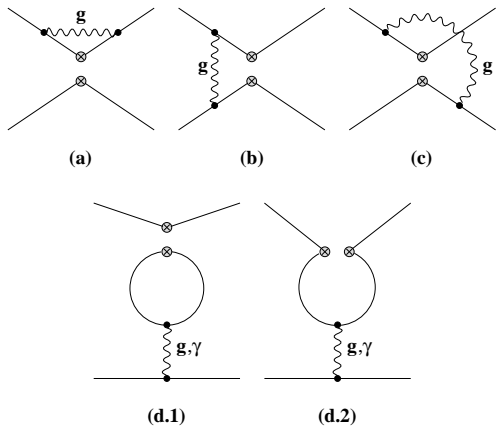}}
\caption{Effective diagrams showing the one-loop operator mixing via 
strong renormalization.}
\label{effdiag}
\end{figure}

In Fig. \ref{effdiag} 
we show examples of how the various coefficients may mix
via QCD renormalization and transmit the properties of the initial
matchings to the other Wilson coefficients at the scale of the low
energy process.

A direct look at the structure of the LO anomalous dimension matrix of the standard 
$\Delta S=1$ effective
quark operators is sufficient to show qualitatively how the initial matching
conditions may feed down to the final values of the
various coefficients.

Here, as a quantitative and model independent test,  
we have varied the NLO (one-loop)
standard-model initial matchings $Y_i \equiv C_i(m_W)$
by factors of ($-1,0,2$) and observed the
effects on the corresponding Wilson coefficients $y_i$ at the scale of 1 GeV.
Our conclusions are the following:
\begin{itemize}
\item
Only the varying of $Y_2$, $Y_6$, $Y_7$ and $Y_9$ leads to 
effects on the low energy $y_i$ larger than a few percents.
($Y_8$ and $Y_{10}$ matchings remain zero at the one-loop level).
\item
Changing the tree level Wilson coefficient $Y_2$, has a proportional
effect on all the gluonic penguin coefficients ($y_{3,4,5,6}$)
and similarly  on
$y_{11,12}$, because of the large additive renormalization induced via
the insertion of $Q_2$ in the penguin like diagrams (d) in Fig. \ref{effdiag}.
The influence on $y_6$ of changing $Y_2$ by a few ten percents
is therefore dramatic for the prediction of \ee. 
On the other hand, one needs a new particle replacing
tree level $W$ exchanges and tree level physics constrains
dramatically these contributions. It is therefore unlikely
to expect sizeable deviations of $Y_2$ from its standard
model value.

\item
Changing $Y_6$ itself
in the range given has no much effect on $y_6$ which
is affected always less than 10\%, and it affects $y_8$ at the 
percent level. Multiplicative renormalization is not the leading
renormalization effect for the gluonic penguins.
\item
Changing $Y_7$ modifies proportionally $y_7$ and $y_8$ 
and may have therefore a dramatic impact on \ee.
\item
Changing $Y_9$ modifies proportionally $y_9$ and $y_{10}$  
and may affects \ee at the few 10\% level via the contribution of $Q_9$.
\end{itemize}

It seems therefore that the most relevant 
potential for new physics effects on \ee resides
in the electroweak penguin sector (see Table \ref{Wcoefs}).
As a matter of fact \cite{Buras:1998ed} have recently shown that bounding
the contribution of the effective $\bar s d Z$ vertex via \ee leads to
the strongest constraints on some rare kaon decays which are governed
by $Z-$penguin diagrams. 

On the other hand, new physics modifications of the standard-model
penguin and box diagrams for $\Delta S = 1,2$ transitions 
affect also the corresponding $\Delta B = 1,2$ amplitudes.
It is therefore likely that in a specific model 
the experimental bounds coming from $B$ physics may indirectly
constrain the deviations
on the electroweak initial matchings within a 
few 10\%~\cite{Nir:1997nj,Fleischer:1997km}.
These bounds would make it hard for new physics to show up in 
visible deviations from the standard \ee prediction. 

The past literature on the subject confirms the  general
conclusion that we reached in the above discussion.
The effect on \ee of charged Higgs particles in the
 two Higgs model has been studied~\cite{Buchalla:1991fu}. The same problem
has also been discussed in the more general framework of
softly broken supersymmetry~\cite{Gabrielli:1995ff}. In both cases 
no significant departures from the
standard model are expected once all bounds are properly implemented.

\section{Conclusions (March 1999)}

On February 24, 1999 the KTeV collaboration has
announced~\footnote{See 
{\tt http://fnphyx-www.fnal.gov/experiments/ktev/ktev.html}} a preliminary
result based on the analysis of 20\% of the data collected which gives
\beq
\Re \: \varepsilon '/\varepsilon =
(28 \pm 3.0\ {\rm (stat)} \pm 2.6\ {\rm (syst)} \pm 1.0\ {\rm (MC\ stat)})
  \times   10^{-4} \, .
\label{KTeV}
\eeq
This result largely deviates from the previous E731 value of \eq{E731}
and sits in the ballpark of the NA31 result (\ref{NA31}).
This value of \ee , if confirmed, signals with high confidence the presence
of direct CP violation in kaon decays, closing successfully a longstanding
and challenging experimental quest. Theoretically, the superweak 
scenario~\cite{Wolfenstein:1964ks} is then excluded as the sole source
of \CP violation. 
 
Averaging (\ref{KTeV}) with (\ref{NA31}) and (\ref{E731}), 
together with the older E731 result 
$\Re \: \varepsilon '/\varepsilon = (32 \pm 30)\times 10^{-4}$,
leads to the value
\beq
\Re \: \varepsilon '/\varepsilon =
(21.8 \pm 3.0) \times 10^{-4} \, .
\label{GrandAve}
\eeq
  
\begin{figure}
\epsfxsize=10cm
\centerline{\epsfbox{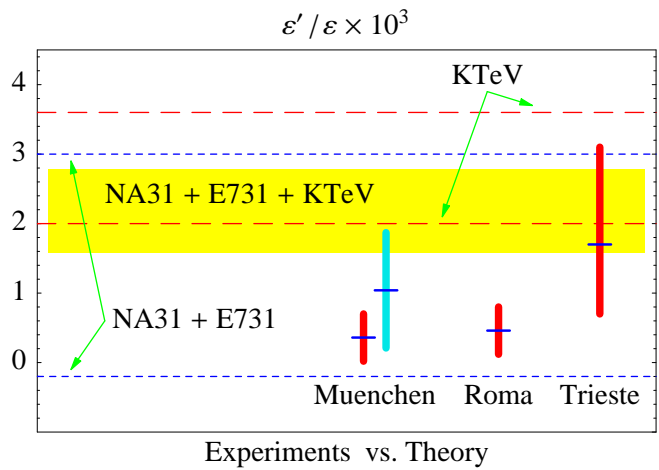}}
\caption{The new $2\sigma$ KTeV result (area enclosed by the 
long-dashed lines) is compared with
the combined $2\sigma$ experimental bounds of
NA31 and E731 (area enclosed by the short-dashed lines).
The combined average in \eq{GrandAve} is shown ($2\sigma$) by the gray band.  
The M\"unchen, Roma and Trieste theoretical estimates for \ee are shown 
with their central values. The M\"unchen and Roma predictions include
gaussian treatment of the input parameters while the uncertainty in
the Trieste estimate corresponds to flat parameter spanning.
The second M\"unchen prediction (light gray)
corresponds to taking a low $m_s$ range (see \eq{lightmsGauss}).
}
\label{fig3}
\end{figure}

In Fig.~\ref{fig3} we update the comparison
between theory and experiment including the new KTeV result.
The light gray area shows the $2\sigma$ range of the  average 
in \eq{GrandAve}. 

The comparison between the present experimental
average and the theoretical predictions shows
a substantial deviation from the Roma and M\"unchen estimates.
Such a disagreement may be reduced by considering a light $m_s$
(see the discussion after \eq{Flat}). The two M\"unchen predictions
shown in Fig.~\ref{fig3}
correspond to $m_s(m_c) = 150\pm 20$ and $m_s(m_c) = 100\pm 20$
(light gray), respectively. 

The Trieste estimate is rather insensitive to $m_s$ since this
parameter enters explicitely only at the NLO in the chiral expansion,
while the value of the quark condensate is determined by the fit
of the $\Delta I = 1/2$ selection rule.
  
Let us recall that the most recent (quenched) lattice estimates
of $m_s$ find $m_s(\mbox{2 GeV}) = 110\pm 23$ 
MeV~\cite{Bhattacharya:1997ht},
corresponding to  $m_s(m_c) = 129\pm 27$ MeV. This range of
$m_s$ is also consistent with recent QCD sum rules 
estimates~\cite{Colangelo:1997uy,Jamin:1997sa}, while
a substantially larger determination of 
$m_s$ is obtained from $\tau$ decays at LEP. A preliminary 
result from the ALEPH collaboration gives $m_s(m_\tau) = 172\pm 31$ 
MeV~\cite{Chen:1998mf}.
In order to assess the theoretical implications 
of the KTeV result it will be important to better 
understand the value of $m_s$, that is currently used to parametrizee
the hadronic matrix elements of the crucial operators
$Q_6$ and $Q_8$ (for a more detailed discussion see~\cite{Keum:1999gw}). 

On the other hand, as we argue in the summary of section VI,
it is premature to take a value of \ee at the $10^{-3}$ level 
as a signal of new physics. In particular, it is worth observing
that:
\begin{itemize}
\item One of the standard model predictions, which via the
$\Delta I = 1/2$ selection obtains all matrix elements, is in 
good agreement with the experimental average in \eq{GrandAve};
\item As it is shown by the VSA and VSA+
toy models, and as it appears from the Trieste calculation, 
the inclusion of NLO order chiral corrections
might alone conspire toward increasing the standard model value obtained
in the present phenomenological and quenched lattice predictions.
\end{itemize}
   
At any rate, efforts on improving all theoretical
estimates are now required. In particular,
a confident assessement of the size of $B_6$ from
lattice will be of crucial relevance.

As discussed in section VI.C,
the uncertainty in all present theoretical estimates may be substantially
reduced by a better determination of $\Im \lambda_t$,
whose error is presently dominated by the uncertainty on $\hat B_K$. 
A precise determination of $\Im \lambda_t$ is expected
in the upcoming years from $B$-physics
at the $B$-factories and at the hadronic colliders. 
In addition, the rare kaon decays $K_L\to \pi^0 \nu\bar\nu$ and
$K^+\to \pi^+ \nu\bar\nu$ provide together a clean probe of 
$\Im \lambda_t$~\cite{Buchalla:1996fp}.
While the latter may be seen at the Brokhaven National Laboratory
within the next year, a new experiment at the same
laboratory has been approved to measure
Br$(K_L\to \pi^0 \nu\bar\nu)$ with a 10\% precision by the year 2005.
This will allow a determination of $\Im \lambda_t$ with a similar accuracy.

In conclusion, the determination of \ee is a
great challenge to both experimentalists and theorists. 
As more precise experimental data become available,
improvements in the theoretical calculations are also expected. The
interplay of the two  will hopefully shed more light on the flavor structure
of the standard model and on some non-perturbative aspects of QCD.

\acknowledgments

SB and MF would like to thank the Physics Department at the University of
Oslo for the hospitality and the financial support during part of
the writing of this review. 
We thank J. Bijnens, A.J. Buras, F.J. Gilman, R. Gupta, G. Martinelli,
E. de Rafael, L. Silvestrini, L. Wolfenstein
and D. Wyler for discussions and comments on the manuscript.


\bibliographystyle{rmp}
\bibliography{ref}

\end{document}